\RequirePackage{fix-cm}
\documentclass[a4paper,12pt,authoryear,fleqn]{article}
\usepackage[applemac]{inputenc}
\usepackage[english]{babel}
\usepackage{natbib}
\usepackage{morefloats}
\usepackage{latexsym}
\usepackage{graphicx}
\usepackage{amsmath}
\allowdisplaybreaks 
\usepackage{amsfonts}
\usepackage{amssymb,tikz}
\usepackage{array}
\usepackage{dsfont}
\usepackage[counterclockwise]{rotating}
\usepackage{multirow}

\usepackage{bigints}
\usepackage{bbm}

\usepackage{grffile}
\graphicspath{{figures/}}
\usepackage{epstopdf}       
\usepackage{setspace}
\renewcommand\arraystretch{1}
\usepackage{sectsty}
\sectionfont{\fontsize{14}{15}\selectfont}
\subsectionfont{\fontsize{13}{15}\selectfont}

\newcommand{\slfrac}[2]{\scalebox{0.55}{$\left.#1\middle/#2\right.$}}

\usepackage{color,soul}
\usepackage{xcolor}
\usepackage{framed}
\colorlet{shadecolor}{yellow}
\definecolor{dgreen}{rgb}{0,0.5,0}
\definecolor{dblue}{rgb}{0,0,0.9}
\definecolor{dred}{rgb}{0.6,0.0,0.1}
\definecolor{dgold}{rgb}{0.5,0.3,0.0}
\definecolor{dvio}{rgb}{0.6,0.3,0.5}
\definecolor{gray}{rgb}{0.5,0.5,0.5}
\definecolor{dbraun}{rgb}{.5,0.2,0}

\usepackage[pdftex,bookmarks=false,pdfpagelabels=false,hidelinks=true,hyperfootnotes=false,hyperfigures = false,hyperindex=false,pageanchor=false,colorlinks=true,citecolor = blue]{hyperref}

\usepackage{ulem}		
\usepackage{textcomp}	

\usepackage{lscape}
\usepackage{afterpage}
\usepackage{capt-of}

\usepackage{tabularx}
\usepackage{longtable}
\usepackage{threeparttable}

\usepackage[ruled,vlined]{algorithm2e}
\SetKwInput{KwInput}{Define}
\SetKwInput{KwOutput}{Step 2}
\SetKwInput{KwData}{Data}
\SetKwInput{KwIn}{Input}

\newtheorem{assum}{Assumption}[section]

\newtheorem{thm}{Theorem}[section]

\newcommand{\bb}{\mbox{\bf b}}

\newcommand{\bZ}{\mbox{\bf Z}}
\newcommand{\bz}{\mbox{\bf z}}
\newcommand{\bx}{\mbox{\bf x}}
\newcommand{\bX}{\mbox{\bf X}}

\newcommand{\wh}{\widehat}

\newcommand{\wtl}{\widetilde}

\newcommand{\EE}{\mathbb{E}}

\newcommand{\btheta}{\boldsymbol{\theta}}

\newcommand{\diag}{\mathrm{diag}}

\newcommand{\1}{\mathbbm{1}}

\makeatletter
\def\env@cases{%
  \let\@ifnextchar\new@ifnextchar
  \left.
  \def\arraystretch{1.2}%
  \array{@{}l@{\,\,}l@{}}%
}%
\makeatother

\makeatletter
\def\BState{\State\hskip-\ALG@thistlm}
\makeatother

\usepackage{xpatch}

\newcommand{\mysetminusD}{\hbox{\tikz{\draw[line width=0.6pt,line cap=round] (3pt,0) -- (0,6pt);}}}
\newcommand{\mysetminusT}{\mysetminusD}
\newcommand{\mysetminusS}{\hbox{\tikz{\draw[line width=0.45pt,line cap=round] (2pt,0) -- (0,4pt);}}}
\newcommand{\mysetminusSS}{\hbox{\tikz{\draw[line width=0.4pt,line cap=round] (1.5pt,0) -- (0,3pt);}}}

\newcommand{\mysetminus}{\mathbin{\mathchoice{\mysetminusD}{\mysetminusT}{\mysetminusS}{\mysetminusSS}}}

\usepackage{verbatim}

\usepackage[draft,colorinlistoftodos,textsize=tiny]{todonotes}
\usepackage{soul}
\makeatletter
\if@todonotes@disabled

\else

\fi
\makeatother

\usepackage{xr}
\usepackage{siunitx}
\usepackage{enumerate}
\usepackage[shortlabels]{enumitem}

\usepackage{subcaption}

\usepackage[left=2cm,right=2cm,top=2cm,bottom=2cm]{geometry}

\usepackage{setspace}
\begin{document}
\setstcolor{red}


\begin{titlepage}
\newgeometry{lmargin=1.20in,rmargin=1.20in,tmargin=0.7in,bmargin=1.20in}

\title{
\begin{singlespace}
\huge Bayesian Bi-level Sparse Group Regressions for Macroeconomic Density Forecasting\thanks{An earlier version of this paper circulated under the title ``Bayesian Bi-level Sparse Group
Regressions for Macroeconomic Forecasting''.}
\end{singlespace}
}

\author{Matteo Mogliani\thanks{
Banque de France - International Macroeconomics Division, 46-1374 DGSEI-DECI-SEMSI, 31 Rue Croix des Petits Champs, 75049 Paris CEDEX 01 (France). Phone: +33(0)142925939. e-mail: \texttt{matteo.mogliani@banque-france.fr}} \and
Anna Simoni\thanks{
CREST, CNRS, \'{E}cole Polytechnique, ENSAE, 5 Avenue Henry Le Chatelier, 91120 Palaiseau (France). Phone: +33(0)170266837. e-mail: \texttt{anna.simoni@polytechnique.edu}}
}




\maketitle

\begin{abstract}
\begin{singlespace}

We propose a Machine Learning approach for optimal macroeconomic density forecasting in a high-dimensional setting where the underlying model exhibits a known group structure. Our approach is general enough to encompass specific forecasting models featuring either many covariates, or unknown nonlinearities, or series sampled at different frequencies. By relying on the novel concept of bi-level sparsity in time-series econometrics, we construct density forecasts based on a prior that induces sparsity both at the group level and within groups. We demonstrate the consistency of both posterior and predictive distributions. We show that the posterior distribution contracts at the minimax-optimal rate and, asymptotically, puts mass on a set that includes the support of the model. Our theory allows for correlation between groups, while predictors in the same group can be characterized by strong covariation as well as common characteristics and patterns. Finite sample performance is illustrated through comprehensive Monte Carlo experiments and a real-data nowcasting exercise of the US GDP growth rate.

\end{singlespace}
\end{abstract}

\begin{singlespace}
\small
\textit{Keywords:} Density forecasts, Data-rich environment, nonlinearities, posterior predictive distribution, Posterior contraction, Spike-and-slab prior\\
\end{singlespace}

\end{titlepage}

\newpage
\newgeometry{lmargin=1.20in,rmargin=1.20in,tmargin=1.20in,bmargin=1.20in}

\setcounter{page}{2}
\setcounter{footnote}{0}
\section{Introduction}\label{sec:Introduction}

Density forecasts of macroeconomic and financial time-series are of crucial importance for policymakers, as they provide an assessment of future (upside or downside) tail risks \citep{GarrattEtAl2003,AdrianEtAl2019,Adams2021,Carriero2024}. Nowadays, researchers dispose of rich datasets, released by official or alternative sources, that potentially contain valuable information for predicting tail events. On the one hand, handling these datasets might be challenging, mainly because of their large dimension (\textit{i.e.} the number of variables is large compared to the number of available observations in the time dimension). On the other hand, the variables in these datasets are (or can be) often organised in groups, a feature that may help the researcher in reducing the dimensionality. The research question of interest is then: how can we optimally exploit group-structures to construct accurate density forecast? 
To the best of our knowledge, there is no clear answer to this question in the literature for general macroeconomic forecasting models. The contribution of this paper is to fill this gap by constructing optimal density forecasts that take advantage of the group-structure for simultaneously \textit{i)} reducing the potentially large dimension in the dataset, and \textit{ii)} evaluating the tail risks probabilities. By detecting, in a data-driven way, the relevant driving factors and groups, our procedure leads to economically interpretable forecasting models and results, which is a crucial feature for decision-taking by policymakers.


Groups might be formed by predictors that present strong covariation, as well as common characteristics and patterns. Some datasets may be organised in groups by the researcher according to economic criteria. For example, real economic activity can be analysed by using a large number of official series arranged in homogeneous blocks of indicators, such as production, employment, consumption, and housing (see \textit{e.g.} \citealp{McCracken2016}). Alternatively, datasets may be organised in a group-structure directly by the data provider. This is the case, for instance, of the Google Search data used in \cite{FerraraSimoniJBES}. In this paper, we widen the relevance of group-structures beyond datasets organised in groups, and we point out that specific forecasting models -- like models with unknown nonlinearities or with covariates observed at mixed frequencies -- might present a group-structure.

To allow for different sources of group-structure, we consider a general model flexible enough to include, among others, \textit{i)} linear regression models with many predictors organised in groups, \textit{ii)} mixed-frequency regression models with nonparametric weighting functions, and \textit{iii)} nonlinear forecasting models as special cases. The nonparametric models \textit{ii)} and \textit{iii)} are suitable for capturing complex relationships between the target variable and the predictors, and the group-structure arises there from generalized Fourier expansions of these relationships.

While the true model is allowed to have an infinite number of elements in each group, we approximate the model by restricting each group to have no more than $g\geq 1$ components. This introduces a bias in finite samples, but this bias vanishes asymptotically as $g$ is allowed to increase with the sample size $T$. The approximate model is assumed to be bi-level sparse: only $s_0^{gr}$ among the $N$ groups are active, and within each of the $s_0^{gr}$ active groups, only a few elements have a non-zero impact on the target variable. This means that some groups and some predictors within an active group can be irrelevant for modelling and forecasting the target variable, conditional on the remaining predictors.

Our Bayesian procedure is based on a hierarchical prior that generates a group structure with bi-level sparsity and  charges only the approximate model. To induce exact bi-level sparsity, we use $(g+1)N$ spike-and-slab priors specified as mixtures between a continuous distribution and a Dirac distribution at zero, the latter inducing exact zeros with non-negligible probability.
%
%
Density forecasts are obtained from the posterior predictive distribution, which is well-known to dominate plug-in predictive distributions when there is prior information. We establish frequentist asymptotic optimality of our Bayesian procedure in a setting where the true data generation process (DGP) has a bi-level sparse group-structure. Importantly, optimality does not require orthogonality between the groups of predictors, such that cross-correlation between covariates in different groups is allowed by our theory. Asymptotic properties are established for the sample size $T$ increasing to infinity, as well as for the number of groups $N$, active groups $s_0^{gr}$, components per groups $g$, and active predictors in the model $s_0$ increasing to infinity with $T$. As a by-product, we provide parameter recovery and point forecasts, as well as the contraction rate of the posterior distribution of the in-sample prediction error. Our posterior contraction rate attains the minimax rate established in \cite{Cai2022IEEE} and \cite{LiZhangYin2024} for bi-level sparse settings.

The bi-level sparse group-structure that we consider in this paper has remarkable advantages over the sparse group-structure considered \textit{e.g.} in \cite{MoglianiSimoni2021}, which only accounts for sparsity among groups but not within groups. First, if the true model is bi-level sparse, then the posterior of \cite{MoglianiSimoni2021} obtained under group sparsity would not put zero mass asymptotically on models with dense groups. Therefore, the true sparsity pattern is not recovered with positive probability, implying a worsening of both selection of relevant driving factors and out-of-sample predictive accuracy. This issue is illustrated in our first Monte Carlo simulation. Second, the contraction rate with bi-level sparsity is faster than the rate with group sparsity if $g\gg \log(N)/\log(T)$ and if $s_0^{gr}g \gg \log(s_0^{gr} g) s_0$. It follows that, in this case, the required sample size for consistency would be large. This is not appealing for macroeconomic forecasting, where the time-series dimension $T$ is usually not large.

Finally, our framework can be easily extended to account for stochastic volatility and ARMA errors, which are important features in several macroeconomic applications and have been proven to improve predictive results \citep{Chan2014,Zhang2020,CCMM2024}. We use this extended setting in an empirical application on density nowcasting the US quarter-on-quarter GDP growth rate.

The paper is structured as follows. In Sections \ref{sec:model} and \ref{sec:priors} we present respectively the model and the prior. Conditional posterior and predictive distributions are provided in Section \ref{sec:MCMC}. The asymptotic properties of our procedure are analysed in Section \ref{sec:Asymptotic_Analysis}, while Monte Carlo experiments are discussed in Section \ref{sec:MCsim}. Section \ref{sec:emp_app} provides an empirical application. Section \ref{sec:conclusions} concludes. Additional theoretical results and all the proofs are reported in the Online Appendix.

\section{The model}\label{sec:model}
\subsection{The sampling model}\label{sec:sampling_model}

Let $y_t$ be the series of interest to be predicted $h$-steps ahead. At each time $t$, a large number of predictors, organized into $N$ groups and denoted by $\bx_{j,t} := \{x_{j,t,i}\}_{i\geq 1}$ for every $j\in\{1,\ldots,N\}$, are available to the forecaster. A group $j$ might be formed, for instance, by indicators belonging to a given sector or category, by lagged values of one predictors or the dependent variable, or by functional transformations of one predictor. Density forecasts of $y_T$, conditional on information available up to $T-h$, are based on the model:
\begin{equation}\label{eq:baseline}
  y_{t} = \sum_{j=1}^N \varphi_j(x_{j,t-h,1}, x_{j,t-h,2},\ldots) + \varepsilon_{t}, \qquad \mathbb{E}\left(\varepsilon_{t}\vert \{\bx_{j,t-h-\ell}\}_{\ell}, \,j=1,\ldots,N\right) = 0
\end{equation}
\noindent for every $h \geq 0$ and $t=1,\ldots,T$. Examples of model \eqref{eq:baseline} will be presented in Section \ref{sec:examples} and an extension to the case where one group contains lagged values of $y_t$ is considered in Appendix A.7. For every $j\in\{1,\ldots, N\}$, $\varphi_j(\cdot)$ denotes a $j$-specific unknown function of $\bx_{j,t-h}$, belonging to a separable Hilbert space $\mathcal{H}_j$ and taking values in $\mathbb{R}$. Depending on the dimension of $\bx_{j,t-h}$, each function $\varphi_j$ might depend on a potentially infinite number of arguments. We assume that $\varepsilon_t$ are independent and identically distributed according to a $\mathcal{N}(0,\sigma^2)$ distribution. Therefore, by introducing the vector $\bx_t := (\bx_{1,t}', \ldots,\bx_{N,t}')'$ of potentially infinite dimension, the matrix $\bX := (\bx_{1-h}, \ldots, \bx_{T-h})'$ with $T$ rows, and the $N$-vectors $\varphi(\bx_t):= (\varphi_1(\bx_{1,t}), \ldots, \varphi_N(\bx_{N,t}))'$ and $\varphi:=(\varphi_1,\ldots,\varphi_N)'$, 
we write the sampling model as:
\begin{equation}\label{eq:sampling:model}
  y_t \vert \bx_{t-h}, \varphi, \sigma^2 \sim \mathcal{N}\left(\sum_{j=1}^N \varphi_{j}(\bx_{j,t-h}),\sigma^2\right), \qquad \forall h \geq 0, \quad\forall t= 1,\ldots,T
\end{equation}
\noindent and the joint sampling distribution of $y := (y_{1},\ldots, y_{T})'$ conditional on $\bX$ is $\prod_{t=1}^T \\\mathcal{N}\left(\sum_{j=1}^N \varphi_{j}(\bx_{j,t-h}),\sigma^2\right)$.\\
\indent For $j=1,\ldots,N$, let $\{z_{j,t-h,i}\}_{i\geq 1}$ be transformations of the elements of $\bx_{j,t-h}$ such that the function $\varphi_j(\bx_{j,t-h})$ writes as $\varphi_j(\bx_{j,t-h}) = \sum_{i=1}^{\infty} \theta_{j,i}z_{j,t-h,i}$. In the remainder of this paper, we shall use this expression for the function $\varphi_j(\cdot)$. 
To reduce the dimension of the model, each function $\varphi_j(\bx_{j,t-h})$ is then approximated by $\sum_{i=1}^g \theta_{j,i} z_{j,t-h,i}$, where $g \geq 1$ is a truncation parameter.
%
We introduce the following notation associated with this approximation: for every $j=1,\ldots,N$, define $\btheta_j := (\theta_{j,1},\ldots,\theta_{j,g})' \in \mathbb{R}^g$, $\btheta := (\btheta_1', \ldots, \btheta_N')'\in\Theta \subset \mathbb{R}^{Ng}$, $\bz_{j,t-h}:= (z_{j,t-h,1},\ldots,z_{j,t-h,g})'$, $\bz_{t} := (\bz_{1,t}^{'},\ldots,\bz_{N,t}^{'})'$ a $(Ng\times 1)$ vector, and $\bZ := (\bz_{1-h},\ldots, \bz_{T-h})'$ a $(T \times Ng)$ matrix. By using this notation, the approximation bias in the mean is:
\begin{multline*}
  \forall t = 1,\ldots,T,\quad B_t(g) \equiv B_{t|t-h}(g) := \mathbb{E}[y_{t}| \bx_{t-h}] - \bz_{t-h}' \btheta\\
  = \sum_{j=1}^N \left(\varphi_j(\bx_{j,t-h}) - \bz_{j,t-h}'\btheta_j\right) = \sum_{j=1}^N \sum_{i>g} \theta_{j,i} z_{j,t-h,i} =: \sum_{j=1}^N B_{t,j}(g),
\end{multline*}
and $B(g):= (B_1(g), \ldots, B_T(g))'$ is a $T$-vector. Therefore, $B_{t,j}(g) := \sum_{i> g} \theta_{j,i}z_{j,t-h,i}$. In this paper we adopt a Bayesian approach and specify a convenient prior for $\btheta$ such that the induced prior for $B_t(g)$ is degenerate at zero (see Section \ref{sec:priors}).

\subsection{The bi-level sparsity structure}\label{sec:sparsity:structure}
Let $(\varphi_{0},\sigma_0^2)$ be the true value of $(\varphi,\sigma^2)$ that generates the data. Under the gaussianity assumption of the error term, the true conditional distribution of $y$ given $\bX$ is $\prod_{t=1}^T\mathcal{N}\left(\sum_{j=1}^N \varphi_{0,j},\sigma_0^2\right)$, with Lebesgue density denoted by $f_0$. The approximation bias of the true sampling mean is denoted by $B_0(g)= (B_{0,1}(g),\ldots, B_{0,T}(g))$ and the true value of $\btheta$ in the approximation of $\varphi_0(\bx_t)$ by $\btheta_{0}:=(\btheta_{0,1}',\ldots, \btheta_{0,N}')'$. The latter is assumed to be bi-level group sparse in the sense explained below.\\
\indent Exact bi-level group sparsity is the feature of the model that guarantees the existence of an approximation $\bz_{t-h}'\btheta_0 \equiv \sum_{j=1}^N \bz_{j,t-h}'\btheta_{0,j}$ to $\sum_{j=1}^N \varphi_{0,j}(\bx_{j,t-h})$ in \eqref{eq:baseline}, with a small number of active groups and of non-zero coefficients for each active group such that the approximation bias $B_{0,t}(g)$ is small relative to the estimation error. To the best of our knowledge, the assumption of exact bi-level sparsity dates back to the recent contributions of \cite{Huang2012}, \cite{Simon2013}, and \cite{Breheny2015}, which propose algorithms for estimation and sparse recovery in cross-section models. Our approach is instead designed for time-series models, with an explicit predictive density motivation and a theoretical foundation.\\
\indent To clarify the concept of exact bi-level sparsity of $\btheta_0$, let $S_{0}^{gr} := \{1 \leq j \leq N; \|\btheta_{0,j}\|_2 > 0\} \subset \{1,2,\ldots,N\}$ be the set of indices of the groups (subvectors) in $\btheta_0$ with at least one nonzero component (active groups), and let $s_{0}^{gr}:=|S_{0}^{gr}|$ denote the cardinality of $S_{0}^{gr}$. If $S_{0}^{gr} \neq \varnothing$, then for every $j \in S_{0}^{gr}$ let $S_{0,j}$ be the set of the indices of the nonzero elements in $\btheta_{0,j}$, such that $S_0 = \bigcup_{j\in S_0^{gr}}|S_{0,j}|$ and $s_{0} := |S_0| =  \sum_{j\in S_{0}^{gr}}|S_{0,j}|$. Remark that $s_{0} \geq 1$ since there is at least one active group under the assumption $S_{0}^{gr} \neq \varnothing$. If $S_{0}^{gr} = \varnothing$, then $s_{0} = 0$. Moreover, $s_0 \equiv s_0(g)$ is a non-decreasing function of $g$. Hence, we say that $\btheta_0$ is $(s_0,s_0^{gr})$-sparse.\\
\indent The next assumption supposes exact bi-level sparsity of $\btheta_0$ and controls the approximation bias $B_{0,t}(g)$ relative to the estimation error. The latter is similar to \citet[Condition ASM]{ChernozhukovBelloniHansen2014}.

\begin{assum}[Exact bi-level sparsity and bias]\label{Ass:4}
  Let $s_0^{gr}, s_0$ be positive integers satisfying $s_0^{gr} \leq N$ and $s_0^{gr} \leq s_0 \leq g s_0^{gr}$. The functions $\{\varphi_{0,j}\}_{j=1,\ldots,N}$ admit the following sparse approximation form: for every $j=1,\ldots,N$ and every $t=1,\ldots,T$, 
  \begin{eqnarray*}
    \varphi_{0,j}(\bx_{j,t-h}) & = & \bz_{j,t-h}' \btheta_{0,j} + B_{0,t,j}(g), \\
    \sum_{j=1}^{N}\1\{\|\btheta_{0,j}\|_2 > 0\} & = & s_0^{gr}, \qquad \sum_{j=1}^N\sum_{i=1}^g\1\{|\theta_{0,ji}|> 0 \} = s_0, \qquad \|B_0(g)\|_2^2 \leq \frac{1}{16}s_0 \sigma_0^2.
  \end{eqnarray*}
\end{assum}

In Section \ref{sec:Asymptotic_Analysis}, we will let $N$, $g$, $s_0$ and $s_0^{gr}$ to increase with $T$. This, together with Assumption \ref{Ass:4}, will allow the size of the approximation model to grow with the sample size $T$. The constant value in the denominator of the upper bound of the approximation bias can be replaced by any constant larger or equal than $16$.

\subsection{Examples}\label{sec:examples}

\subsubsection{Linear regression models with grouped predictors}\label{sec:grouped_pred}
Many datasets used for macroeconomic forecasting display a large panel of real and financial indicators. Theses series can be often organized in homogeneous groups by following either economic information or statistical procedures. For example, a nominal group for different price indicators, an output group for supply and production indicators, a financial group for interest rates and stock prices, etc.
In this paper, we propose a procedure for density forecasting that takes into account the group structure in large datasets.

We suppose that each group of covariates contains at most $g$ elements. Let $y_t$ be the target variable to be predicted $h$-steps ahead,
$j\in\{1,\ldots,N\}$ be the group index, and $\bx_{j,t}$ be the $g$-vector of variables in each group $j$. Then, by assuming a linear model: $\forall t=1,\ldots,T$ and $h>0$,
\begin{equation}\label{eq:group_reg}
  y_{t} = \sum_{j=1}^N \bx_{j,t-h}'\btheta_j + \varepsilon_{t}, \qquad \mathbb{E}\left(\varepsilon_{t}| \{\bx_{j,t - h - \ell}\}_{\ell},\,j=1,\ldots,N\right) = 0
\end{equation}
which can be cast in model \eqref{eq:baseline} with $\bz_{j,t} = \bx_{j,t}$, $\varphi_j(\bx_{j,t-h}) = \bx_{j,t-h}'\btheta_j$ and zero approximation bias.

\subsubsection{Nonparametric mixed-frequency regression models}\label{sec:midas}

Consider a low-frequency variable $y_{t}^{L}$, where $t=1,\ldots,T$ indexes the low frequency time unit, and consider $(N-1)$ high-frequency variables $x_{j,t}^H$ for $j=1,\ldots, N-1$. By denoting with $m$ the number of times the higher sampling frequency appears in the low-frequency time unit $t$, then  $t - k/m$ denotes the $k$-th past high frequency period for $k = 0,1,2,\ldots$. To simplify the notation, we set the same $m$  for all the groups, but our framework can accommodate the case with a $m_j$ specific to each group.
Let us define the high-frequency lag operator $L^{\slfrac{1}{m}}$, such that $L^{\slfrac{1}{m}}x_{j,t}^{H}=x_{j,t-\slfrac{1}{m}}^{H}$. Further, let $h=0,1/m,2/m,3/m,\dots$ be an (arbitrary) forecast horizon. For given orders $p_y, p_x\geq 0$, the general Mixed Data Sampling (MIDAS) regression model can be written as follows: $\forall t=1,\ldots,T$,
\begin{equation}\label{eq:MIDAS}
y_{t}^{L} = \sum_{j=1}^{N}\Psi_j(L^{1/m})x_{j,t-h}^{H} + \varepsilon_{t}^{L}, \qquad \EE\left(\varepsilon_{t}^L \vert \{x_{j,t - h - \ell}\}_{\ell, \,j=1,\ldots,N}\right) = 0,
\end{equation}
where $\Psi_j(L^{1/m})$ is the high-frequency lag polynomial
\begin{equation}\label{eq:HF_lg_poly}
\Psi_j(L^{1/m})=\sum_{u=0}^{p_x}\psi_j(u)L^{u/m},\qquad \psi_j(\cdot):\mathbb{R}_+ \rightarrow \mathbb{R}.
\end{equation}
This model can be cast in model \eqref{eq:baseline} with $N$ groups, where $\varphi_j(\bx_{j,t-h}) = \Psi_j(L^{1/m})x_{j,t-h}^{H}$ for every $j=1,\ldots,N$ with $\bx_{j,t-h} = (x_{j,t-h}^H, \ldots, x_{j,t-h - p_x/m}^H)^{\prime}$.\footnote{Model \eqref{eq:MIDAS} can be generalized to accommodate groups of indicators sampled at the same frequency as the target variable $y_{t}$.} 
Previous literature has considered different parameterizations of the weighting function $\psi_j(u)$ in \eqref{eq:HF_lg_poly}. Parametric specifications have been considered, for instance, by \citet{Foroni2015} and \citet{Ghysels2007}. 
\citet{MoglianiSimoni2021} use (non-orthogonalized) algebraic power polynomials which correspond to restricted Almon lag polynomials, while \cite{Babii2022} use shifted orthogonal polynomials such as Jacobi and Legendre. 
With the method developed in the present paper, we can allow $\psi_j(\cdot)$ to belong to a separable Hilbert space $\mathcal{H}_j$ with a countable orthonormal basis $\{\phi_1(\cdot), \phi_2(\cdot),\ldots\}$ and construct density forecasts for MIDAS models. Therefore, for any $\psi_j(\cdot)\in\mathcal{H}_j$, we can write
\begin{equation}\label{eq:weighting:function:non:truncated}
\psi_j(u) = \sum_{i=1}^{\infty}\theta_{ji}\phi_i(u),
\end{equation}
\noindent where $\theta_{ji} := \langle \psi_j,\phi_i\rangle$ is the $i$-th Fourier coefficient. Hence,
\begin{displaymath}
  \Psi_j(L^{1/m}) x_{j,t-h}^H = \sum_{u=0}^{p_x}\sum_{i=1}^{\infty}\theta_{ji}\phi_i(u) x_{j, t-h-u/m}^H = \sum_{i=1}^{\infty} \theta_{ji}\Phi_i^{\prime} \bx_{j, t-h} = \varphi_j(\bx_{j, t-h}),
\end{displaymath}
\noindent where $\Phi_i := (\phi_i(0),\phi_i(1), \ldots, \phi_i(p_x))^{\prime}$. By cutting the sum in $i$ at some $g > 0$, we get
$  \varphi_j(\bx_{j, t-h}) = \Psi_j(L^{1/m})x_{j,t-h}^H = \sum_{i=1}^{g}\theta_{ji}\Phi_i^{\prime} \bx_{j, t-h} + B_{t,j}(g)$,
which yields a mixed-frequency regression model with approximated high-frequency polynomial $\sum_{i=1}^{g}\theta_{ji}\Phi_i^{\prime} \bx_{j, t-h}$, $z_{j,t-h,i} = \Phi_i^{\prime} \bx_{j, t-h}$, $\bz_{j,t-h} = (\bx_{j, t-h}^{\prime}\Phi_1, \ldots, \bx_{j, t-h}^{\prime}\Phi_g)'$, and an approximation bias at time $t$ given by $B_{t,j}(g) = \sum_{i > g}\theta_{ji}\Phi_i^{\prime} \bx_{j, t-h}$. 

\subsubsection{Nonlinear predictive regression models with interaction effects}\label{sec:nonlinear}
In many settings, it may be desirable to account for possible nonlinearities in the conditional mean function of the target variable $y_t$. Model \eqref{eq:baseline} can accommodate different types of nonlinearities. The simplest case is when, given $p$ covariates, the effect of each covariate on $y_t$ can be separated in $p$ nonlinear functions and interaction effects are taken into account by using an additive partially linear model: $\forall t=1,\ldots,T$ and $h>0$,
 \begin{equation}\label{eq:nonlinear:interaction}
  y_{t} = \sum_{j=1}^p \varphi_j(x_{j,t-h})+ z_{t-h}'\btheta_{p+1} + \varepsilon_{t}, \qquad \mathbb{E}\left(\varepsilon_{t} \vert \{x_{j,t-h-\ell}\}_{\ell,\, j= 1, \ldots, N}\right) = 0,
\end{equation}
\noindent where $z_{t-h} := \{x_{j,t-h}x_{k,t-h}\}_{k>j}$ is a $p(p-1)/2$-vector that includes the interactions between covariates.
Here, we have $N=p+1$ groups, with the last group having a very high number of components (of the order $p^2$), and $\varphi_j$ is a function of only one covariate taking values in a separable Hilbert space $\mathcal{H}_j$ with a countable orthonormal basis $\{\phi_{j,1}(\cdot), \phi_{j,2}(\cdot),\ldots\}$: $\varphi_j\in\mathcal{H}_j$. Then, for every $j\in\{1,\ldots,p\}$, $\varphi_j(x_{j,t-h}) = \sum_{i=1}^{\infty}\theta_{ji}\phi_{j,i}(x_{j,t-h})$, where $\theta_{ji} := \langle \varphi_j,\phi_{j,i}\rangle$ is the $i$-th Fourier coefficient and $\langle \cdot,\cdot\rangle$ denotes the inner product in $\mathcal{H}_j$. For some $g>0$ we can approximate $\varphi_j(x_{j,t-h})$ as
\begin{equation*}
\varphi_j(x_{j,t-h}) \approx \sum_{i=1}^{g}\theta_{ji}\phi_{ji}(x_{j,t-h}) = : \bz_{j,t-h}'\btheta_j,
\end{equation*}
\noindent which yields an approximation bias at time $t$ given by $B_{t,j}(g) = \sum_{i > g}\theta_{ji}\phi_{ji}(x_{j,t-h})$.
%
%
Alternatively, we can replace $z_{t-h}'\btheta_{p+1}$ with $\sum_{j=1}^p z_{j,t-h}'\btheta_{p+j}$ in \eqref{eq:nonlinear:interaction}, where $z_{j,t-h} := \{x_{j,t-h}x_{k,t-h}\}_{k\neq j}$ is a $(p-1)$-vector. The total number of groups would hence be $N=2p$, with the last $p$ groups of dimension $p-1$ each, which is much lower than $p^2$. It follows that the choice of the group structure may imply a trade-off between the number of groups and the number of components of each group.

\section{The BSGS-SS prior}\label{sec:priors}
\indent With the approximation model in Section \ref{sec:sampling_model} and the Assumption \ref{Ass:4} in mind, we elicit a prior that puts all its mass on the approximation $\bz_{t-h}'\btheta$, conditional on $\bz_{t-h}$, and that induces sparsity at the group level and within groups. Let us first define, for every group $j=1,\ldots,N$, the following quantities: $\btheta_{j} = \mathbf{V}_j^{1/2}\bb_{j}$, $\bb_{j} := (b_{j1},\ldots,b_{jg})'$, $\mathbf{V}_j^{1/2} := \diag(v_{j1},\ldots, v_{j g})$ and $v_{ji} \geq 0$ for $i=1,\ldots, g$.\\
\indent We treat the truncation parameter $g$ as deterministic and, under Assumption \ref{Ass:4}, it may depend on $s_0$. Our proposed Bayesian Sparse Group Selection with Spike-and-Slab prior (BSGS-SS henceforth) is specified as: $\forall j=1,\ldots, N $,
\begin{align}
  \theta_{ji} & \sim \delta_0,\qquad \forall i > g, \label{eq:prior:dirac}\\
  \bb_{j}|g,\pi_0 & \stackrel{ind.}{\sim} (1 - \pi_0)\mathcal{N}_{g}(0,I_{g}) + \pi_0 \delta_0(\bb_{j}), \label{eq:prior:b}\\
  v_{ji}|\pi_1,\tau_j & \stackrel{ind.}{\sim} (1 - \pi_1) \mathcal{N}^+(0,\tau_j^2) + \pi_1 \delta_0(v_{ji}),\qquad \quad i=1,\ldots, g, \label{eq:prior:v}\\
  \tau_j & \stackrel{ind.}{\sim} \mathcal{G} \left(\lambda_{0},\lambda_{1,j}\right), \qquad \lambda_{0} = \frac{1}{2},\\
  \pi_0 & \sim \mathcal{B}(c_0,d_0), \qquad \pi_1 \sim \mathcal{B}(c_1,d_1),\qquad \sigma^2 \sim \mathcal{G}^{-1} (a_{0},a_{1}), \label{eq:prior:sigma}
\end{align}
\noindent where $\mathcal{N}^+(0,\tau_{j}^{2})$ denotes a $\mathcal{N}(0,\tau_{j}^{2})$ distribution truncated below at zero, $\delta_0(\cdot)$ denotes a Dirac distribution at zero, $\mathcal{G}$ (resp. $\mathcal{G}^{-1}$) denotes the Gamma (resp. Inverse-Gamma) distribution, $\lambda_{1,j}$ in the prior for $\tau_j$ is the scale parameter, and $\mathcal{B}$ denotes a Beta distribution.
The conditional priors for $\bb_{j}|g,\pi_0$ and for $v_{ji}|\pi_1,\tau_j^2$ are specified as hard spike-and-slab priors. A similar prior has been considered in \citet[Section 3.2]{Xu2015}. However, a subtle difference here stems from the non-conjugate Gamma prior for the standard deviation hyperparameter $\tau_j$. We prove in Section \ref{sec:Asymptotic_Analysis} that this specification is crucial to ensure good asymptotic properties for the corresponding posterior. Compared with the priors in \citealp{ChenChuYuanWu2016}, the proposed prior is computationally convenient as we can simulate from the posterior distribution by using a relatively simple Gibbs sampler with one Metropolis-Hastings step.\\
\indent By integrating out $\{v_{ji}\}_{i=1}^g$, the induced prior on $\theta_{ji}|\pi_0,\pi_1,\tau_j$ is: $\forall j=1,\ldots,N$,
\begin{multline}
  \theta_{ji}|\pi_0,\pi_1,\tau_j,g \sim (1 - \pi_1)(1 - \pi_0)(1 - \pi_1^g)f_{\theta_{ji}}(\theta_{ji}|\tau_j) \, \\
  + \, \delta_0(\theta_{ji})\left(\pi_1 + [\pi_1^g(\pi_0 - 1) - \pi_0](\pi_1 - 1)\right), \qquad \forall i=1,\ldots, g,\label{eq:prior:theta}
\end{multline}
where the Lebesgue density $f_{\theta_{ji}}$ is upper bounded by the Lebesgue density of a $\mathcal{G}(1/2,\tau_j)$ distribution. Its explicit expression is given in the Online Appendix A.1, together with a proof of this result. In the appendix we also discuss the conditional prior on the random function $\varphi_j$, given $\{v_{ji}\}_{i=1}^g$ and $g$, induced by the prior \eqref{eq:prior:dirac}-\eqref{eq:prior:b}.\\
\section{The BSGS-SS predictive density and the MCMC algorithm}\label{sec:MCMC}

Let $\Pi(\cdot|y,\mathbf{X})$ denote the posterior of the model's parameters $(\varphi,\sigma^2)$ obtained from the sampling model \eqref{eq:sampling:model} and the BSGS-SS prior in Section \ref{sec:priors}. We construct density forecasts based on the posterior predictive density of $y_{\tau}|\bx_{\tau-h}$, denoted by $\wh{f}(y_{\tau}|\bx_{\tau - h},y,\bX)$ and conditional on a new value of the covariates $\bx_{\tau - h}$, for $\tau>T$:
\begin{equation}\label{eq:pred_dens}
  \wh{f}(y_{\tau}|\bx_{\tau - h},y,\bX):= \int f_0(y_{\tau}|\bx_{\tau - h},\varphi,\sigma^2) \Pi(d\varphi,d\sigma^2|y,\bX)
\end{equation}
\noindent where $f_0(y_{\tau}|\bx_{\tau - h},\varphi,\sigma^2) $ denotes the Lebesgue density associated with the sampling distribution in \eqref{eq:sampling:model}.
%
Draws from the predictive distribution \eqref{eq:pred_dens} can be obtained directly from the MCMC sampler detailed in Algorithm \ref{algorithm:1}, where
we denote by ``rest'' the conditioning parameters and data in each conditional distribution. 

The hyperparameters of the priors on $\pi_{0}$ and $\pi_{1}$ must be chosen very carefully as they control the overall amount of between-group and within-group prior sparsity. In this paper, we propose to select $c_0$ and $c_1$ by using a data-driven approach based on the Deviance Information Criterion (DIC; \citealp{Spiegelhalter2002}), defined as:
\begin{equation}\label{eq:DIC}
  \textrm{DIC}(\mathbf{c}|y,\bX) := -4 \mathbb{E}_{\boldsymbol\btheta,\sigma^2}\big[\log f(y\vert \boldsymbol\btheta,\sigma^2,\bX)\vert y,\bX, \mathbf{c} \big] + 2\log f(y\vert \widehat{\btheta}_{\mathbf{c}},\widehat\sigma_{\mathbf{c}}^2,\bX),
\end{equation}
where $(\btheta^{\prime}_{\mathbf{c}},\sigma^{2}_{\mathbf{c}})^{\prime}$ denotes the model parameters for a given set of hyperparameters $\mathbf{c}=(c_{0},c_{1})$, and $(\widehat{\btheta}_{\mathbf{c}},\widehat\sigma_{\mathbf{c}}^2)$ is a point estimate of $(\btheta^{\prime}_{\mathbf{c}},\sigma^{2}_{\mathbf{c}})^{\prime}$.\footnote{The first part of \eqref{eq:DIC} is the posterior mean deviance, which is here estimated by averaging the log-likelihood function, $\log f(y\vert \boldsymbol\btheta,\sigma^2,\bX)$, over the posterior draws of $(\btheta^{\prime}_{\mathbf{c}},\sigma^{2}_{\mathbf{c}})^{\prime}$. $(\widehat{\btheta}_{\mathbf{c}},\widehat\sigma_{\mathbf{c}}^2)$ is computed using the posterior median for $\btheta$ and the posterior mean for $\sigma^2$.}
This metrics of fit presents the advantage of being easily available from the MCMC output, with no additional costs in terms of computational burden.\footnote{We select the $c_{0}$ and $c_{1}$ that minimize the DIC by performing a random search on a fine 2-dimensional grid, with lower-bounds set as described in (A.1.3). In empirical applications, $s_{0}^{gr}$ and $s_{0}$ are unknown, but we recommend to tune them at some arbitrary values such that the inequalities in (A.1.3) are satisfied for low values of $c_{0}$ and $c_{1}$. Therefore, the price to pay for not knowing $s_{0}^{gr}$ and $s_{0}$ is simply the burden of searching on a larger grid.}

  \begin{algorithm}[!t]
  \caption{MCMC sampling scheme}\label{algorithm:1}
\footnotesize{
\KwInput{$\eta^{2}_{ji} := \left(b_{ji}^2 \mathbf{Z}_{ji}^{\prime}\mathbf{Z}_{ji}\sigma^{-2} + \tau^{-2}_{j}\right)^{-1}$, $\nu_{ji} := \sigma^{-2}\eta^{2}_{ji} b_{ji}\mathbf{Z}_{ji}^{\prime}\left[y - \mathbf{Z}_{j\mysetminus i}\boldsymbol\theta_{j\mysetminus i}\right]$, $\boldsymbol\Sigma_{j} := (\sigma^{-2} \mathbf{V}_{j}^{1/2}\mathbf{Z}_{j}^{\prime}\mathbf{Z}_{j}\mathbf{V}_{j}^{1/2} + \mathbf{I}_{m_{j}})^{-1}$,
$\boldsymbol\mu_{j} := \sigma^{-2}\boldsymbol\Sigma_{j} \mathbf{V}_{j}^{1/2}\mathbf{Z}_{j}^{\prime}\left[y - \mathbf{Z}_{\mysetminus j}\boldsymbol\theta_{\mysetminus j}\right]$, where $\mathbf{Z}_{j\mysetminus i}\boldsymbol\theta_{j\mysetminus i}=\mathbf{Z}_{\mysetminus j}\boldsymbol\theta_{\mysetminus j} + \sum_{k\neq i}b_{jk} v_{jk}\mathbf{Z}_{jk}$ and $\mathbf{Z}_{\mysetminus j}\boldsymbol\theta_{\mysetminus j}=\sum_{\kappa\neq j}b_{\kappa i} v_{\kappa i}\mathbf{Z}_{\kappa i}$. }
\KwData{$y$, $\mathbf{Z}$.}
\KwIn{$\mathbf{b}^{(0)}$, $\sigma^{2(0)}$, $\{\mathbf{V}_j^{1/2,(0)}\}_{j=1,\ldots,N}$, $\pi_{0}^{(0)}$, $\pi_{1}^{(0)}$, $\boldsymbol\tau^{(0)}$,$a_{0}$, $a_{1}$, $c_{0}$, $c_{1}$, $d_{0}$, $d_{1}$.}
\For{$\ell = 1,\ldots, MC$}{
\begin{itemize}
  \item[(1)] Sample $\sigma^{2(\ell)}\vert \textrm{rest} \sim \mathcal{G}^{-1}\left(\frac{T}{2}+a_{0},\frac{1}{2}\Vert y-\mathbf{Z}\boldsymbol{\theta}^{(\ell-1)}\Vert_{2}^{2} + a_{1}\right)$.
  \item[(2)] $\forall j=1,\dots,N$ and $i=1,\dots,g$:
    \begin{enumerate}[label=(2-\arabic*)]
      \item Update $\eta^{2(\ell)}_{ji}$ and $\nu_{ji}^{(\ell)}$, and sample $\gamma_{1,ji}^{(\ell)}|rest \sim \textrm{Bernoulli}(\widetilde{\pi}_{1,ji}^{(\ell)})$, where
        $$\widetilde{\pi}_{1,ji}^{(\ell)} = \frac{\pi_{1}^{(\ell-1)}}{\pi_{1}^{(\ell-1)}+2(1-\pi_{1}^{(\ell-1)})\frac{\eta_{ji}^{(\ell)}}{\tau_{j}^{(\ell)}}\exp\left(\frac{\nu^{2(\ell)}_{ji}}{2\eta^{2(\ell)}_{ji}}\right)\Phi\left(\frac{\nu_{ji}^{(\ell)}}{\eta_{ji}^{(\ell)}}\right)}$$
        and $\Phi(\cdot)$ denotes the cdf of the Normal distribution.
      \item Sample $v_{ji}^{(\ell)} \vert \textrm{rest} \sim (1-\gamma_{1,ji}^{(\ell)})\mathcal{N}^{+}\left(\nu_{ji}^{(\ell)},\eta_{ji}^{2(\ell)}\right)  + \gamma_{1,ji}^{(\ell)}\delta_{0}(v_{ji}^{(\ell)})$ and set $\mathbf{V}_j^{1/2,(\ell)} = \diag(v_{j1}^{(\ell)},\ldots, v_{jg}^{(\ell)})$.
    \end{enumerate}
    \item[(3)] $\forall j=1,\dots,N$:
    \begin{enumerate}[label=(3-\arabic*)]
      \item Update $\boldsymbol\mu_{j}^{(\ell)}$, $\boldsymbol\Sigma_{j}^{(\ell)}$, and sample $\gamma_{0,j}^{(\ell)}|rest \sim \textrm{Bernoulli}(\widetilde{\pi}_{0,j}^{(\ell)})$, where
        $$\widetilde{\pi}_{0,j}^{(\ell)} = \frac{\pi_{0}^{(\ell-1)}}{\pi_{0}^{(\ell-1)}+(1-\pi_{0}^{(\ell-1)})\vert\boldsymbol\Sigma_{j}^{(\ell)}\vert^{\frac{1}{2}}\exp\left(\frac{1}{2}\boldsymbol\mu_{j}^{(\ell)\prime}(\boldsymbol\Sigma_{j}^{(\ell)})^{-1}\boldsymbol\mu_{j}^{(\ell)}\right)}.$$
      \item Sample $\mathbf{b}_{j}^{(\ell)} \vert \textrm{rest} \sim  (1-\gamma_{0,j}^{(\ell)})\mathcal{N}\left(\boldsymbol\mu_{j},\boldsymbol\Sigma_{j}\right)  + \gamma_{0,j}^{(\ell)}\delta_{0}(\mathbf{b}_{j})$.
      \item Update $\btheta_{j}^{(\ell)} = \mathbf{V}_j^{1/2,(\ell)}\bb_{j}^{(\ell)}$.
    \end{enumerate}
    \item[(4)] $\forall j=1,\dots,N$, sample $\tau_{j}^{(\ell)}$ conditional on $\mathbf{V}_{j},\lambda_{1,j}$ using a Metropolis-Hastings step with proposal density an exponential distribution. Draw $\tau_{j}^{new}\sim\mathcal{E}xp(\widetilde{\lambda})$, where $\widetilde{\lambda}^{-1}=\tau_{j}^{(\ell - 1)}$. Set $\tau_j^{(\ell)} = \tau_j^{new}$ with probability $\min(1,R)$, where
        \begin{multline*}
          R = \left(\frac{\tau_{j}^{(\ell - 1)}}{\tau_{j}^{new}}\right)^{\left(\xi_{j}+\frac{3}{2}\right)}\exp\left\{-\frac{1}{2}\sum_{i;v_{ji}^{(\ell)}>0}v_{ji}^{2(\ell)}\left[\frac{1}{(\tau_{j}^{new})^{2}}-\frac{1}{\tau_{j}^{2(\ell - 1)}}\right] - \frac{(\tau_{j}^{new}-\tau_{j}^{(\ell - 1)})}{\lambda_{1,j}}\right\}\\
          \times \exp\left\{-\frac{\tau_{j}^{(\ell - 1)}}{\tau_{j}^{new}} + \frac{\tau_{j}^{new}}{\tau_{j}^{(\ell - 1)}}\right\}
        \end{multline*}
        and $\xi_{j}:=\sharp\{i; v_{ji}^{(\ell)} > 0\}$. 
    \item[(5)] Sample
      $\pi_{0}^{(\ell)} \vert \textrm{rest} \sim \mathcal{B}\left(N-\sum_{j=1}^N\gamma_{0,j}^{(\ell - 1)}+c_{0},\sum_{j=1}^N\gamma_{0,j}^{(\ell - 1)}+d_{0}\right)$ and $\pi_{1}^{(\ell)} \vert \textrm{rest} \sim \mathcal{B}\left(Ng-\sum_{j=1}^N\sum_{i=1}^{g}\gamma_{1,ji}^{(\ell - 1)}+c_{1},\sum_{j=1}^N\sum_{i=1}^{g}\gamma_{1,ji}^{(\ell - 1)}+d_{1} \right)$.
\end{itemize}

%
}}
\end{algorithm}


\normalsize

\subsection{Extending the model to a general error process}\label{sec:SVOtARMA}

One of the main advantage of the proposed prior is its flexibility on the treatment of the error process. Unlike other shrinkage priors (Group Lasso, Sparse Group Lasso), it is quite straightforward to extend the prior in Section \ref{sec:priors} to account, for instance, for stochastic volatility and ARMA errors. The relevance of these features in macroeconomic applications has been extensively stressed by the literature (\text{e.g.}, \citealp{Clark2011,Chan2014,Carriero2019}). Further, recent works \citep{CCMM2024,LenzaPrimiceri2022} have suggested that time-varying volatility could attenuate inferential issues (\textit{e.g.}, explosive patterns and widening uncertainty for forecasts and Impulse Response Functions) when the estimation sample includes extreme observations such as those recorded during the COVID-19 pandemic.\\
\indent We then modify the homoskedastic model in \eqref{eq:baseline} to account for a general error process as follows:
\begin{align}
  y_{t} &= \sum_{j=1}^N \varphi_j(x_{j,t-h,1}, x_{j,t-h,2},\ldots) + \varepsilon_{t} \\ 
  \phi_{\varepsilon}(L)\varepsilon_{t} &= \psi_{u}(L)u_{t},\qquad  u_{t} \sim t_{\nu}\left(0,\omega_{t}\exp(\zeta_{t})\right), \label{eq:baseline_svotarma_arma} 
\end{align}
\noindent where we assume a stationary and invertible ARMA($p$,$q$) specification for the errors $\varepsilon_{t}$ and heavy-tails innovations $u_t$, the latter featuring time-varying log-volatility $\zeta_{t}$ and outliers component $\omega_{t}$ in the variance.
Since the Student-$t$ distribution $t_{\nu}$ in \eqref{eq:baseline_svotarma_arma} can be expressed as a scale mixture of Normal distributions with scale term $\tau_{t} \sim \mathcal{G}^{-1}(0.5\nu,0.5\nu)$, then,
conditional on $(\tau_{t}, \omega_{t}, \zeta_{t})$, the innovations have distribution $u_{t} \sim \mathcal{N}\left(0,\tau_{t}\omega_{t}\exp(\zeta_{t})\right)$.
The log-volatility $\zeta_{t}$ is assumed to evolve according to a stationary AR(1) process:
$\zeta_{t} \sim \mathcal{N}\left(\mu_{\zeta} + \phi_{\zeta}(\zeta_{t-1}-\mu_{\zeta}),\sigma^{2}_{\zeta}\right),$
with initial conditions $\zeta_{0} \sim \mathcal{N}\left(\mu_{\zeta},\sigma^{2}_{\zeta}/(1-\phi_{\zeta}^{2})\right)$ and $\phi_{\zeta}$ restricted to lie in the stationary region. While the scale term $\tau_{t}$ is intended to capture frequent, although limited, jumps in the volatility, the additional scale term $\omega_{t}$ is instead intended to capture infrequent, but extreme, jumps. Following \citet{Stock2016} and \citet{CCMM2024}, the prior for $\omega_{t}$ has a two-part distribution, depending on whether an observation at period $t$ is considered either regular or outlier: $\omega_t =1$ with probability $(1-p_{\omega})$ and $\omega_t \sim \mathcal{U}(2,\bar{\omega})$ with probability $p_{\omega}$ for a constant $\bar{\omega} > 2$,
where $\mathcal{U}(2,\bar{\omega})$ denotes the Uniform distribution with support $(2,\bar{\omega})$.

The implementation of these features implies only minor modifications to the MCMC Algorithm \ref{algorithm:1} (see the Online Appendix A.3 and Section \ref{sec:emp_app} for an application). Conditional posteriors for stochastic volatility and ARMA parameters are the same as in \citet{Kastner2014}, \cite{Chan2013}, and \citet{Zhang2020}, and we refer the reader to these contributions for more details.

\section{Theoretical properties}\label{sec:Asymptotic_Analysis}

This section provides the theoretical validation of our procedure. In Section \ref{ss:density_forecast_consistency} we establish consistency of the predictive density for density forecast. As a by-product, in Section \ref{ss:posterior_consistency} we demonstrate consistency of the posterior distribution of the model's parameters $(\varphi,\sigma^2)$, denoted by $\Pi(\cdot|y,\mathbf{X})$. We adopt a frequentist point of view, in the sense that we admit the existence of a true value $(\varphi_{0},\sigma_0^2)$ that generates the data. We denote by $\mathbb{E}_0[\cdot]$ the expectation taken with respect to the true data distribution $P_0 = \prod_{t=1}^T \mathcal{N} \left(\sum_{j=1}^N \varphi_{0,j}(\bx_{j,t-h}),\sigma_0^2\right)$ of $y$ conditional on $(\varphi_{0},\sigma_0^2)$ and $\bX$ with Lebesgue density denoted by $f_0$. All the analysis is conditional on $\bZ$. In the following, we denote by $\|\bZ\|_{op}$ the operator norm and $\|\bZ\|_o:=\max\{\|\bZ_j\|_{op}; 1\leq j\leq N\}$, where $\bZ_j$ is the $(T \times g)$-submatrix of $\bZ$ made of all the rows and the columns corresponding to the indices in the $j$-th group. Moreover,  $\|\cdot\|_2$ denotes the Euclidean norm. Denote
$$\epsilon := \max\{\sqrt{s_0^{gr} \log(N)/T},\sqrt{s_0\log(T)/T},\sqrt{s_0\log(s_0^{gr}g)/T}\}$$
the rate of contraction of the posterior distribution of $\varphi$. The first term corresponds to the complexity of identifying $s_0^{gr}$ non-zero groups, while the third term corresponds to the complexity of estimating $s_0$ non-zero elements of a parameter distributed in $s_0^{gr}$ known groups. In absence of group-structure, the maximum between these two rates corresponds to the rate for recovery of sparse vectors over $\ell_0$-balls (see \textit{e.g.} \citealp{RaskuttiWainwright2011}). This case can be obtained by setting either only one group (\textit{i.e.} $N=1$ and $g = p$) or a number of groups equal to the number of $p$ parameters (\textit{i.e.} $N=p$ and $g=1$).

For two sequences $a_T$ and $b_T$ and two constants $c_1,c_2>0$ we write $a_T \asymp b_T$ if $c_1 b_T\leq a_T \leq c_2 b_T$. By using this notation, if $\log(s_0^{gr} g) \asymp \log(s_0^{gr} g/s_0)$, $\log(N) \asymp \log(N/s_0^{gr})$, and $s_0\log(T) \leq \min\{s_0^{gr}\log(N),s_0\log(s_0^{gr} g)\}$, then $\epsilon$ corresponds to the minimax rate for recovering $\varphi$ given in \cite{Cai2022IEEE} and in \cite{LiZhangYin2024}. The conditions $\log(s_0^{gr} g) \asymp \log(s_0^{gr} g/s_0)$ and $\log(N) \asymp \log(N/s_0^{gr})$ guarantee a sparse setting, while the condition $s_0\log(T) \leq \min\{s_0^{gr}\log(N),s_0\log(s_0^{gr} g)\}$ guarantees a high-dimensional setting. If $s_0 = s_0^{gr} g$, which corresponds to the case with no sparsity within the groups, then the last condition corresponds to Assumption 4.1\textit{(ii)} in \citet{MoglianiSimoni2021}.\\
\indent The following assumption restricts some of the parameters of the model.
\begin{assum}\label{Ass:1}
  (i) For positive and bounded constants $\underline{\sigma}^2$, $\overline{\sigma}^2$, $0 < \underline{\sigma}^2 \leq \sigma_0^2 \leq \overline{\sigma}^2<\infty$; (ii) $\max\{N, T\} \leq e^{s_0^{gr} g}$; (iii) $\max_{j\in S_0^{gr}}\max_{i\in S_{0,j}}|\theta_{0,ji}| \leq \log(s_0^{gr} g)$.
\end{assum}

Assumption \ref{Ass:1} \textit{(i)} excludes degenerate cases by restricting the model variance. Assumption \ref{Ass:1} \textit{(ii)} restricts the rate at which the number of groups can increase compared to $T$, $s_0^{gr}$ and $g$. It is violated for instance if $N>\max\{T,\exp\{s_0^{gr} g\}\}$. Assumption \ref{Ass:1} \textit{(iii)} restricts the growth rate of the largest component of $\btheta_0$.

\indent The next two assumptions concern the hyperparameters of the prior. Assumption \ref{Ass:2} allows $\lambda_{1,j}$ (the hyperparameter of the prior for $\tau_j$) to increase or decrease with $T$, $g$, $s_0^{gr}$ and $|S_{0,j}|$. It rules out a $\lambda_{1,j}$ decreasing too fast to zero or increasing too fast to infinity. In practice, any positive constant can be taken as a value for $\lambda_{1,j}$ and its choice depends on the desired tightness of the prior for $\tau_j$.

\begin{assum}\label{Ass:2}
  Assume that $\sqrt{T}/(\|\mathbf{Z}\|_o \min\{\log(g s_0^{gr}),\log(T)\}) < C$ with probability $1$ for some constant $C>0$. The scale parameters $\lambda_{1,j}$ are allowed to change with $T$ and belong to the range: for every $j=1,\ldots,N$,
  $$\max\left\{\frac{1}{|S_{0,j}|},\frac{\sqrt{T}}{\|\mathbf{Z}\|_o }\right\}\underline{c} \leq \lambda_{1,j} \leq \lambda_{\max} \leq \overline{C} \min\{\log(s_0^{gr} g),\log(T))\}$$
  for two positive constants $1 <\underline{c}< \overline{C}<\infty$, where $\lambda_{\max} := \max\{\lambda_{1,j};\, j\leq N\}$.
\end{assum}
For the following assumption we introduce the function $(u)_+ := \max\{u,0\}$.
\begin{assum}\label{Ass:3}
  There exist positive constants $\kappa_0$ and $\kappa_1$ such that the hyper-parameters $c_0, d_0, c_1, d_1$ of the Beta priors for $\pi_0$ and $\pi_1$ satisfy: \textit{(i)} for every $N>1$, $(d_0+j-1)/(c_0 + N - j) \leq \kappa_0 j/[N^{u_0}(N - j + 1)]$ for every $ u_0 \in (\log(2) / \log(N) ,  \leq s_0^{gr}]$ and
  $\forall j\in\{1,\ldots,N\}\subseteq\mathbb{N}$, \textit{(ii)} for every $s_0^{gr}g > 1$, $(d_1+j - 1)/(c_1 + Ng - j) \leq \kappa_1 j/[(s_0^{gr}g)^{u_1}(Ng - j + 1)]$ for every $ u_1 \in (\log(2)/\log(s_0^{gr}g), s_0]$ and $\forall j\in\{1,\ldots,g\}\subseteq\mathbb{N}$, \textit{(iii)} for a positive constant $C_{cd}$,
  $$\max\left\{c_1, d_0\log(N + c_0 + d_0), d_1\log(s_0^{gr} g)\right\} \leq C_{cd}T\epsilon^2;$$
  \textit{(iv)} $c_1 \geq \lambda_{1,j}^2 g \frac{3d_1}{4} - d_1$.
\end{assum}

Assumption \ref{Ass:3} \textit{(i)} and \textit{(ii)} require that $c_0$ and $c_1$ increase together with $N$, $s_0^{gr}$ and $g$ and control their rate. To satisfy the assumption, if $d_0$ and $d_1$ are chosen equal to one and if $\kappa_0$ and $\kappa_1$ are chosen such that $\kappa_0 < N^{u_0}$ and $\kappa_1 < (s_0^{gr}g)^{u_1}$, then $c_0$ and $c_1$ have to be at least of the order of $1 - N +N^{u_0+1}/\kappa_0$ and $(s_0^{gr} g)^{u_1} Ng/\kappa_1 - Ng + 1$, respectively, for $u_0$, $u_1$ in the range of values given in the assumption and up to a constant. On the other hand, if $d_0$ and $d_1$ are chosen to increase with $T$, then $c_0$ and $c_1$ have to increase faster than $d_0N^{u_0 + 1}$ and $d_1(s_0^{gr} g)^{u_1}$. In practice, one can set the constants $\kappa_0$ and $\kappa_1$ at very small values, as long as they are fixed and do not increase with $T$. Assumptions \ref{Ass:3} \textit{(i)} correspond to those in \citet[Assumption 1]{Castillo2015} for the special case of a Beta prior. We provide in Appendix A.2 a deeper analysis of this assumption. 

\subsection{Consistency of the predictive distribution}\label{ss:density_forecast_consistency}
We now provide a validation of our procedure for density forecasts. 
The next theorem states that, for every $\tau>T$, $\wh{f}(y_{\tau}|\bx_{\tau - h},y,\bX)$ in \eqref{eq:pred_dens} converges to the Lebesgue density $f_0(y_{\tau}|\bx_{\tau - h},\varphi_0,\sigma_0^2)$ of the true distribution $\mathcal{N} \left(\sum_{j=1}^N \varphi_{0,j}(\bx_{j,\tau-h}),\sigma_0^2\right)$ with respect to the Hellinger distance denoted by $d_H(\cdot,\cdot)$.
\begin{thm}\label{thm3:consistency:OOS}
For every $\tau>T$, let $f_0(y_{\tau}|\bx_{\tau - h},\varphi_0,\sigma_0^2)$ denote the Lebesgue density function of the true conditional distribution $P_0$ of $y_{\tau}$ given $\bx_{\tau-h}$ evaluated at $y_{\tau}$. Suppose Assumptions \ref{Ass:4} and \ref{Ass:1} - \ref{Ass:3} hold and let $\epsilon \rightarrow 0$  as $T\rightarrow \infty$. Suppose that $\left|B_{0,\tau-h}(g)\right|^2\lesssim \epsilon^2$, and that there exist positive constants $\kappa_{\ell}$ and $\kappa_z$ such that for all $j\geq 1$ and some $\beta > 2$,
\begin{multline}
  \Pi\left(\left.e^{4jM_2\epsilon^2} < \frac{\sigma^2 + \sigma_0^2}{2\min\{\sigma_0^2,\sigma^2\}} \leq e^{4(j+1)M_2\epsilon^2}]  \right|y,\bX\right) \leq j^{-\beta},\\
  \Pi\left(\left.\frac{M_3\epsilon^2}{\kappa_\ell \kappa_z}j<\|\btheta - \btheta_0\|_2^2 \leq \frac{M_3\epsilon^2}{\kappa_\ell \kappa_z}(j+1)\right|y,\bX\right)\leq j^{-\beta}.\label{thm:OOS:conditions}
\end{multline}
\noindent Then,
\begin{displaymath}
  \lim_{T\rightarrow\infty} P_0\left(\left.d_H^2(f_0(y_{\tau}|\bx_{\tau - h},\varphi_0,\sigma_0^2), \wh{f}(y_{\tau}|\bx_{\tau - h},y,\bX))\leq\varepsilon\right|\bX,\bx_{\tau - h}\right) = 1
\end{displaymath}
\noindent for all $\varepsilon > 0$ and all $\bx_{\tau-h}$ such that $\sum_{j\in S^{gr}}\sum_{i\in S_{j}}z_{j,\tau-h,i}^2< C$ for some constant $C>0$, $\forall S^{gr}\subseteq\{1,\ldots,N\}$ such that $|S^{gr}| \leq s_0^{gr}+ T\epsilon^2/\log(N)$, and $\forall S_j\subseteq\{1,\ldots,g\}$ such that $\sum_{j\in S_0^{gr}}|S_j| \leq s_0 + T\epsilon^2/\log(s_0^{gr} g)$. 
\end{thm}
We know from (A.6.18) and Theorem A.5.1 in the Online Appendix that the probabilities in \eqref{thm:OOS:conditions} go to zero. The assumption of the theorem is then just an assumption about the rate at which these posterior probabilities go to zero. We refer to \cite{HoffmannRousseauSH2015} for primitive conditions for this assumption.

\subsection{Consistency of the posterior distribution}\label{ss:posterior_consistency}
We establish here the posterior consistency uniformly over the infinite dimensional set $\mathcal{F}(s_0,s_0^{gr};\bZ)$ defined as:
\begin{multline*}
  \mathcal{F}(s_0,s_0^{gr};\bZ) := \Big\{(\varphi,\sigma^2); \|B(g)\|_2^2 \leq \frac{s_0 \sigma^2}{16},
  s_{\btheta}^{gr} \leq s_0^{gr}, \, s_{\btheta} \leq s_0, \, \\
  \max_{j\in S_{\btheta}^{gr}}\max_{i\in S_{\btheta,j}}|\theta_{ji}| \leq \log(s_0^{gr} g),\textrm{ and }\sigma^2 \in[\underline{\sigma}^2, \overline{\sigma}^2]\Big\}
\end{multline*}
\noindent for $s_0^{gr}, s_0 \in \mathbb{N}_+$ satisfying $s_0^{gr} \leq N$ and $s_0^{gr} \leq s_0 \leq g\, s_0^{gr}$. The next theorem establishes the in-sample consistency of our BSGS-SS procedure. It shows that the posterior contracts in a neighborhood of the true value defined by the in-sample prediction error. We use the notation $\varphi_{j}^{(T)}(\bX) := (\varphi_j(\bx_{j,-h+1}), \ldots, \varphi_j(\bx_{j,T-h}))'$ for $j\in\{1,\ldots,N\}$ and $\mathcal{H} = \mathcal{H}_1\times \cdots \times \mathcal{H}_N$.
\begin{thm}\label{thm:1}
  Suppose Assumptions \ref{Ass:4} and \ref{Ass:1} - \ref{Ass:3} hold and let $\epsilon \rightarrow 0$ as $T\rightarrow \infty$. Then, for a sufficiently large constant $M >0$: as $T\rightarrow \infty$,
  \begin{displaymath}
    \sup_{(\varphi_0,\sigma_0^2)\in\mathcal{F}(s_0,s_0^{gr};\bZ)}\mathbb{E}_0 \left[\Pi\left(\left.\varphi\in \mathcal{H}; \left\|\sum_{j=1}^N \left(\varphi_j^{(T)}(\bX) - \varphi_{0,j}^{(T)}(\bX)\right)\right\|_2^2 \leq M T\epsilon^2\right|y,\mathbf{X}\right)\right] \rightarrow 1.
  \end{displaymath}
\end{thm}
A similar result is obtained for the case with lagged values of the dependent variable among the covariates (see the Online Appendix A.7).
The required sample size to achieve convergence to zero of the in-sample prediction error in Theorem \ref{thm:1} is $T> C\max\{s_0^{gr}\log(N), s_0 \log(s_0^{gr} g), s_0\log(T)\}$ for some positive constant $C$. In the grouped predictors example of Section \ref{sec:grouped_pred}, the norm in Theorem \ref{thm:1} is the inner product weighted by the Gram matrix $\bX'\bX$, that is $\|\bX(\btheta - \btheta_0)\|_2^2$.
In the mixed-frequency example of Section \ref{sec:midas}, the norm is:
$$\left\|\sum_{j=1}^N \left(\varphi_j^{(T)}(\bX) - \varphi_{0,j}^{(T)}(\bX)\right)\right\|_2^2 = \sum_{t=1}^T\left(\sum_{j=1}^N \sum_{i=1}^{\infty}\left(\theta_{j,i} - \theta_{0,ji}\right)\Phi_i^{\prime} \bx_{j, t-h}\right)^2.$$

In addition to the entire function $\varphi$, it is worth focusing on the coefficients $\btheta$. The consistency of the marginal posterior of $\btheta$ is established in Theorem A.5.1 in the Online Appendix. The contraction rate for $\btheta$ coincides with the rate obtained in \citet[Corollary 1]{LiZhangYin2024} for a linear regression model.\\
\indent We now look at the posterior mean of $\varphi(\bx_t)$ as a possible point estimator for $\varphi_0(\bx_t)$. Convergence of a Bayesian point estimator towards the true $\varphi_0$ is in general not implied by the result of Theorem \ref{thm:1} if the loss function is not bounded. The next theorem analyzes the asymptotic behaviour of the posterior mean in terms of the Euclidean loss function $\ell_h(\wtl\varphi,\varphi) := \frac{1}{T}\sum_{t=1}^T\left(\sum_{j=1}^N\left[\wtl\varphi_j(\bx_{t - h}) - \varphi_j(\bx_{t - h})\right]\right)^2$, which is not uniformly bounded if the parameter space is not compact.

 \begin{thm}\label{thm:4:3}
   Let us consider the posterior mean estimator $\wh\varphi(\bx_t) := \mathbb{E}[\varphi(\bx_t) \vert y,\bX]$, for $t \leq T$ and assume that as $T\rightarrow\infty$, $\mathbb{E}_0\left[\Pi\left(\left.\ell_h(\varphi,\varphi_0) > M j \epsilon^2\right|y,\bX\right)\right] \leq C j^{-\beta}$ for every $j \geq 1$ and for some $\beta> 2$ and some constant $C>0$. Then under the assumptions of Theorem \ref{thm:1},
   $$\mathbb{E}_0\left[\frac{1}{T}\sum_{t=1}^T\left(\sum_{j=1}^N\left[\wh\varphi_j(\bx_{t - h}) - \varphi_{0,j}(\bx_{t - h})\right]\right)^2\right] \rightarrow 0\qquad \textrm{as } T\rightarrow\infty.$$
 \end{thm}

\noindent The result of the theorem holds under the assumption that the posterior of the complement of a ball centered on $\varphi_0$, with radius proportional to the contraction rate $\epsilon^2$, becomes sufficiently small for large $T$. Results of this type have been established for instance in \cite{HoffmannRousseauSH2015}.

\section{Monte Carlo experiments}\label{sec:MCsim}
We evaluate the finite sample performance of our procedure through two Monte Carlo experiments. First, we consider a DGP featuring grouped predictors, as in the example described in Section \ref{sec:grouped_pred}. Second, we consider a DGP featuring mixed-frequency data, as in the example described in Section \ref{sec:midas}. Simulations are based on 200 Monte Carlo iterations, each featuring 60'000 MCMC sweeps (with a burn-in sample of 10'000 sweeps and a chain-thinning parameter set at 5).

\subsection{Experiment 1: DGP with grouped predictors}\label{sec:MCsim_Exp1}
In this experiment, the DGP involves $Ng=\{100,300\}$ predictors (sampled at the same frequency as the target variable) structured in $N=\{5,10,20\}$ groups. The group active set has cardinality $s_{0}^{gr}=\{1,5,10\}$ and we fix $|S_{0,j}|=1$ for all simulations, such that $s_0 = s_0^{gr}$. The elements of $S_{0}^{gr}$ and $S_{0,j}$ are fixed realizations of random draws without replacement from $\{1,\dots,N\}$ and $\{1,\dots,g\}$, respectively. The number of in-sample observations is fixed at $T=200$, and the out-of-sample at $T_{oos}=50$.
Simulated data are obtained from the following process:
\begin{align}
y_{t} &= \alpha+\beta_{y} y_{t-1} + \sum_{j=1}^N \sum_{i=1}^{g} z_{j,t,i} \theta_{ji}+\varepsilon_{t}, \label{eq:DGP1_baseline}\\[1.5ex]
z_{j,t,i} &= \rho_{z} z_{j,t-1,i} + \epsilon_{j,t,i}, \\[1.5ex]
\left( \begin{array}{c}
\varepsilon_{t}\\
\boldsymbol\epsilon_{t}
\end{array} \right) &\thicksim \textrm{i.i.d.} ~ \mathcal{N}
\left[ \left( \begin{array}{c}
0 \\
\mathbf{0}
\end{array} \right),
\left( \begin{array}{c c}
\sigma^{2} & \mathbf{0} \\
\mathbf{0} & \boldsymbol\Sigma_{\epsilon} \\
\end{array} \right)
\right],\label{sim:1:errors}
\end{align}
where $\boldsymbol\epsilon_{t}:=(\epsilon_{1,t,1},\ldots,\epsilon_{N,t,g})'$, $\boldsymbol\Sigma_{\epsilon}=\mathcal{S}_{\epsilon}\mathcal{R}_{\epsilon}\mathcal{S}_{\epsilon}$ is a block-diagonal matrix, $\mathcal{S}_{\epsilon}$ a $(Ng \times Ng)$ diagonal matrix with elements $\sigma_{\epsilon}$, and $\mathcal{R}_{\epsilon}$ a block-diagonal Toeplitz correlation matrix, with $N$ blocks each of size $(g \times g)$ and featuring diagonal elements equal to one and off-diagonal elements $\rho_{\epsilon}^{|i-i^{\prime}|}$ for all $i\neq i^{\prime}$. 
The group structure is therefore given by the structure of this covariance matrix.
We set $\sigma=0.50$ and $\rho_{\epsilon}=0.50$, and we calibrate $\sigma_{\epsilon}$ such that the noise-to-signal ratio of \eqref{eq:DGP1_baseline} is $\text{NSR}=0.20$ for all the simulations. The coefficients of the active variables is set to $|\theta_{ji}|=0.5$, for each $j\in S_{0}^{gr}$ and $i\in S_{0,j}$. The sign of the coefficients is a fixed realization of random draws with replacement from $\{-1,1\}$. Finally, we set $\alpha=0.2$, $\beta_{y}=0.3$, and $\rho_{z}=0.9$.

Simulation results are reported in Table \ref{t:mc_results3}, where we focus on both selection and density forecast performance. The former is evaluated by computing the Matthews correlation coefficient at the group level (MCC$_{N}$), which measures the overall quality of the groups classification, while the latter is evaluated by the means of the Continuously Ranked Probability Score (CRPS), averaged over the out-of-sample observations and expressed in relative terms with respect to an AR(1) benchmark. The results for our prior (BSGS-SS) are then compared to those obtained from the Bayesian Sparse Group Lasso (BSGL) prior of \citet{Xu2015} and the Bayesian Adaptive Group Lasso with Spike-and-Slab prior (BAGL-SS) prior of \citet{MoglianiSimoni2021}. The values reported in the table denote average outcomes across Monte Carlo iterations and their bootstrap standard errors (in parentheses).

The results for our BSGS-SS prior suggest that true group active set can be correctly selected with extremely high probability (MCC$_{N}$ close to one) and point to accurate density forecasts (CRPS less than one). This holds irrespective of the number of groups $N$ and for DGPs involving a very sparse environment (\textit{i.e.} when $s_{0}^{gr}$ and $s_{0}$ are small compared to $N$ and $g$). However, consistently with the contraction rates derived in Section \ref{sec:Asymptotic_Analysis}, selection and predictive accuracy tend to deteriorate in less sparse DGPs.
Compared to the alternative priors considered, the results also point to a substantial outperformance of our BSGS-SS prior, in particular for highly sparse DGPs. Focusing on density forecasts, three main findings can be remarked. First, the predictive accuracy of our model is almost unaffected by the increase in the total number of predictors, while both the BSGL and the BAGL-SS priors display a fairly strong deterioration of their forecasting performance. Second, as mentioned above, our model fails to significantly outperform the AR(1) benchmark in a less sparse environment, but it still significantly outperforms the competing models. Third, the predictive performance of single-level group sparsity priors, such as the BAGL-SS, may quickly deteriorate in presence of complex sparsity structures (such as bi-level sparsity).

\begin{table}[t!]
\footnotesize
\caption{Monte Carlo Experiment 1 - selection and density forecast accuracy}
\centering
\begin{tabular}{c|c|c|c|cc|cc|cc}
\hline\hline
\multicolumn{4}{c}{} & \multicolumn{2}{c}{BSGS-SS} & \multicolumn{2}{c}{BSGL} & \multicolumn{2}{c}{BAGL-SS}\\
\hline
$Ng$ & $N$ & $g$ & $s_{0}^{gr}$ & MCC$_{N}$ & CRPS & MCC$_{N}$ & CRPS & MCC$_{N}$ & CRPS \\
\hline
\multirow{10}{*}{100} & \multirow{ 2}{*}{5} & \multirow{ 2}{*}{20} & \multirow{ 2}{*}{1} & 0.98 & 0.70 & 0.82 & 0.75 & 0.86 & 0.79 \\[-0.5ex]
			       &  &  &  & \scriptsize(0.01) & \scriptsize(0.01) & \scriptsize(0.01) & \scriptsize(0.01) & \scriptsize(0.02) & \scriptsize(0.01) \\
\cline{2-10}
			       & \multirow{ 4}{*}{10} & \multirow{ 4}{*}{10} & \multirow{ 2}{*}{1} & 0.99 & 0.69 & 0.88 & 0.72 & 0.84 & 0.75 \\[-0.5ex]
			       &  &  &  & \scriptsize(0.01) & \scriptsize(0.01) & \scriptsize(0.01) & \scriptsize(0.01) & \scriptsize(0.02) & \scriptsize(0.01) \\[0.5ex]
			       &  &  & \multirow{ 2}{*}{5} & 0.98 & 0.73 & 0.52 & 0.93 & 0.77 & 1.05 \\[-0.5ex]
			       &  &  &  & \scriptsize(0.01) & \scriptsize(0.01) & \scriptsize(0.01) & \scriptsize(0.01) & \scriptsize(0.01) & \scriptsize(0.01) \\
\cline{2-10}
       			       & \multirow{4}{*}{20} & \multirow{4}{*}{5} & \multirow{ 2}{*}{5} & 1.00 & 0.72 & 0.68 & 0.87 & 0.87 & 0.90 \\[-0.5ex]
			       &  &  &  & \scriptsize(0.01) & \scriptsize(0.01) & \scriptsize(0.02) & \scriptsize(0.01) & \scriptsize(0.01) & \scriptsize(0.02) \\[0.5ex]
       			       &  &  & \multirow{ 2}{*}{10} & 0.56 & 0.93 & 0.23 & 1.00 & 0.58 & 1.14 \\[-0.5ex]
			       &  &  &  & \scriptsize(0.01) & \scriptsize(0.02) & \scriptsize(0.01) & \scriptsize(0.01) & \scriptsize(0.01) & \scriptsize(0.02) \\
			       			
\cline{1-10}
			
\multirow{10}{*}{300} & \multirow{ 2}{*}{5} & \multirow{ 2}{*}{60} & \multirow{ 2}{*}{1} & 1.00 & 0.71 & 0.70 & 0.88 & 0.84 & 1.03 \\[-0.5ex]
			       &  &  &  & \scriptsize(0.01) & \scriptsize(0.01) & \scriptsize(0.01) & \scriptsize(0.01) & \scriptsize(0.02) & \scriptsize(0.02) \\
\cline{2-10}
			       & \multirow{4}{*}{10} & \multirow{4}{*}{30} & \multirow{ 2}{*}{1} & 0.99 & 0.70 & 0.76 & 0.81 & 0.91 & 0.85 \\[-0.5ex]
			       &  &  &  & \scriptsize(0.01) & \scriptsize(0.01) & \scriptsize(0.01) & \scriptsize(0.01) & \scriptsize(0.01) & \scriptsize(0.01) \\[0.5ex]
                                &  &  & \multirow{ 2}{*}{5} & 0.97 & 0.73 & 0.37 & 1.12 & 0.54 & 1.31 \\[-0.5ex]
			       &  &  &  & \scriptsize(0.01) & \scriptsize(0.01) & \scriptsize(0.01) & \scriptsize(0.02) & \scriptsize(0.02) & \scriptsize(0.03) \\
\cline{2-10}
       			       & \multirow{4}{*}{20} & \multirow{4}{*}{15} & \multirow{ 2}{*}{5} & 0.96 & 0.74 & 0.49 & 1.04 & 0.71 & 1.15 \\[-0.5ex]
			       &  &  &  & \scriptsize(0.01) & \scriptsize(0.01) & \scriptsize(0.01) & \scriptsize(0.01) & \scriptsize(0.01) & \scriptsize(0.02) \\[0.5ex]
       			       &  &  & \multirow{ 2}{*}{10} & 0.33 & 1.00 & 0.22 & 1.10 & 0.38 & 1.32 \\[-0.5ex]
			       &  &  &  & \scriptsize(0.01) & \scriptsize(0.01) & \scriptsize(0.01) & \scriptsize(0.01) & \scriptsize(0.01) & \scriptsize(0.03) \\
\hline\hline
\multicolumn{10}{l}{\scriptsize \parbox[t]{12.0cm}{Notes: $T=200$, $s_{0,j}=1$, $s_{0}=s_{0}^{gr}$. MCC$_N$ and CRPS denote respectively the Matthews Correlation Coefficient at the group level and the Continuously Ranked Probability Score, the latter expressed in relative terms with respect to the AR(1) benchmark. Average values over 200 Monte Carlo simulations. Bootstrap standard errors of Monte Carlo averages in parentheses.}}
\end{tabular}
\label{t:mc_results3}
\end{table}

Additional simulation results and robustness checks are reported in the Online Appendix A.4. In particular, we show that in a very sparse environment, the mean squared error (MSE) is extremely low (with almost zero bias) and the true active set (both at group and variables level) is correctly selected with extremely high probability (according to both the True Positive Rate and the Matthews correlation coefficient). Further, we show that the performance of our model is robust to different correlation structures in $\boldsymbol\Sigma_{\epsilon}$ and alternative error distributions (\textit{e.g.} Skew-Normal distribution). Overall, the results are very close to those reported in Table \ref{t:mc_results3}, irrespective of the values for $Ng$, $N$, and $s_0^{gr}$, although they tend to deteriorate with a substantial increase in the noise-to-signal ratio.\footnote{We perform the following exercises: \textit{i)} we increase the within-groups correlation parameter by setting $\rho_{\epsilon}=0.75$; \textit{ii)} we allow for between-groups correlation by letting the Toeplitz correlation matrix $\mathcal{R}_{\epsilon}$ to be a full matrix, \textit{i.e.} $\mathcal{R}_{\epsilon}=\rho_{\epsilon}^{|\iota-\iota^{\prime}|}$ for all $\iota\neq \iota^{\prime}$, with $\iota=1,\dots,Ng$; \textit{iii)} we allow for asymmetric shocks in the DGP by letting $\varepsilon_{t}$ to follow a $\text{Skew-}\mathcal{N}(\xi,\omega^{2},\alpha)$ distribution \citep{Azzalini2014}, with shape parameter $\alpha=-5$, and location and scale parameters $(\xi,\omega)$ calibrated such that $\varepsilon_{t}$ has mean zero and variance $\sigma^{2}$; \textit{iv)} we increase the noise-to-signal ratio to 0.5.}

\begin{figure}[!t]
\caption{MCC$_g$ for the BSGS-SS and the BAGL-SS priors across DGPs} \label{fig:MCCg}
\begin{subfigure}[b]{0.50\textwidth}
	\centering
		  \includegraphics[width=\linewidth]{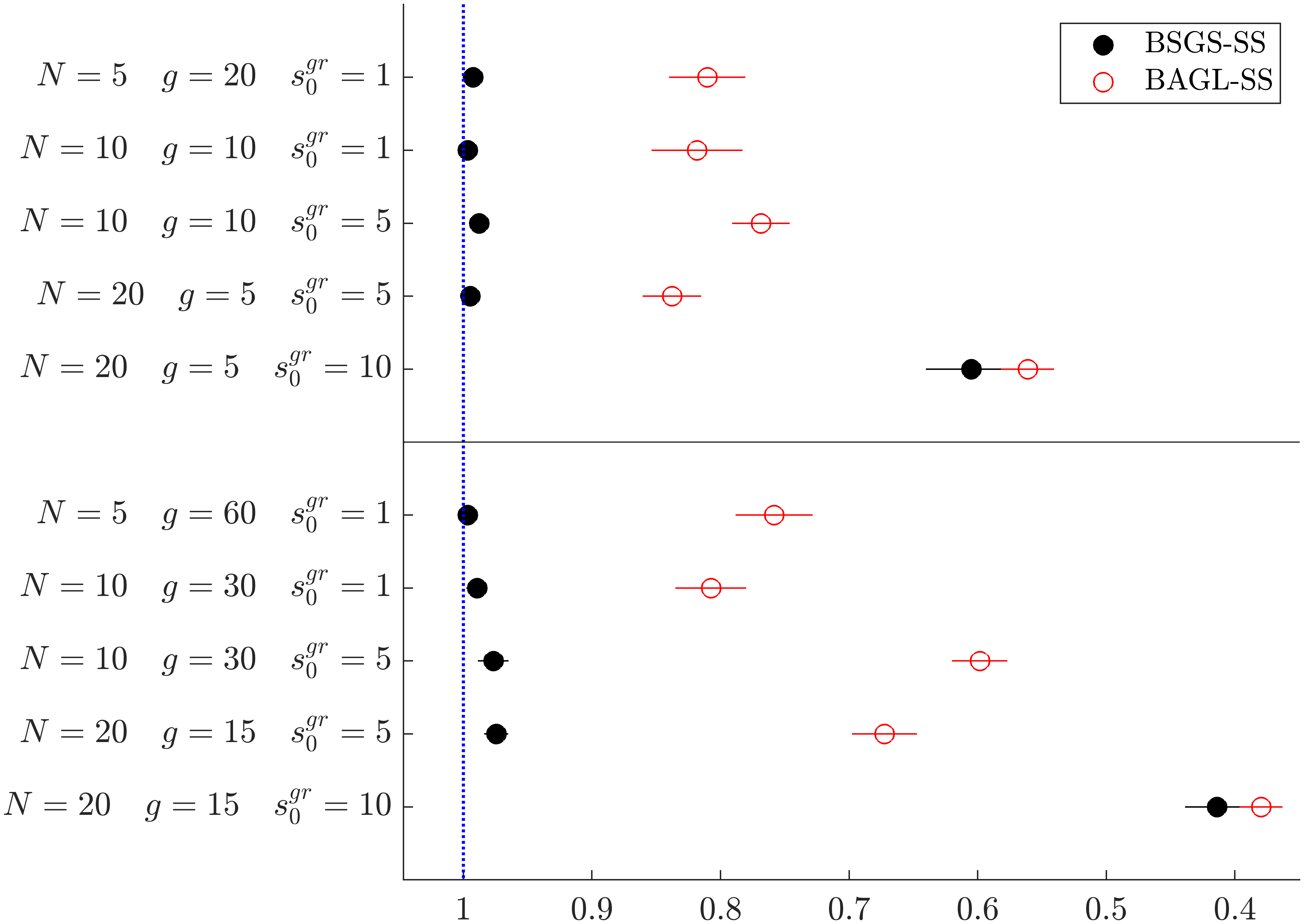}
    \vspace{-0.5cm}
    \caption{Baseline}
\end{subfigure}
\begin{subfigure}[b]{0.50\textwidth}
	\centering
		  \includegraphics[width=\linewidth]{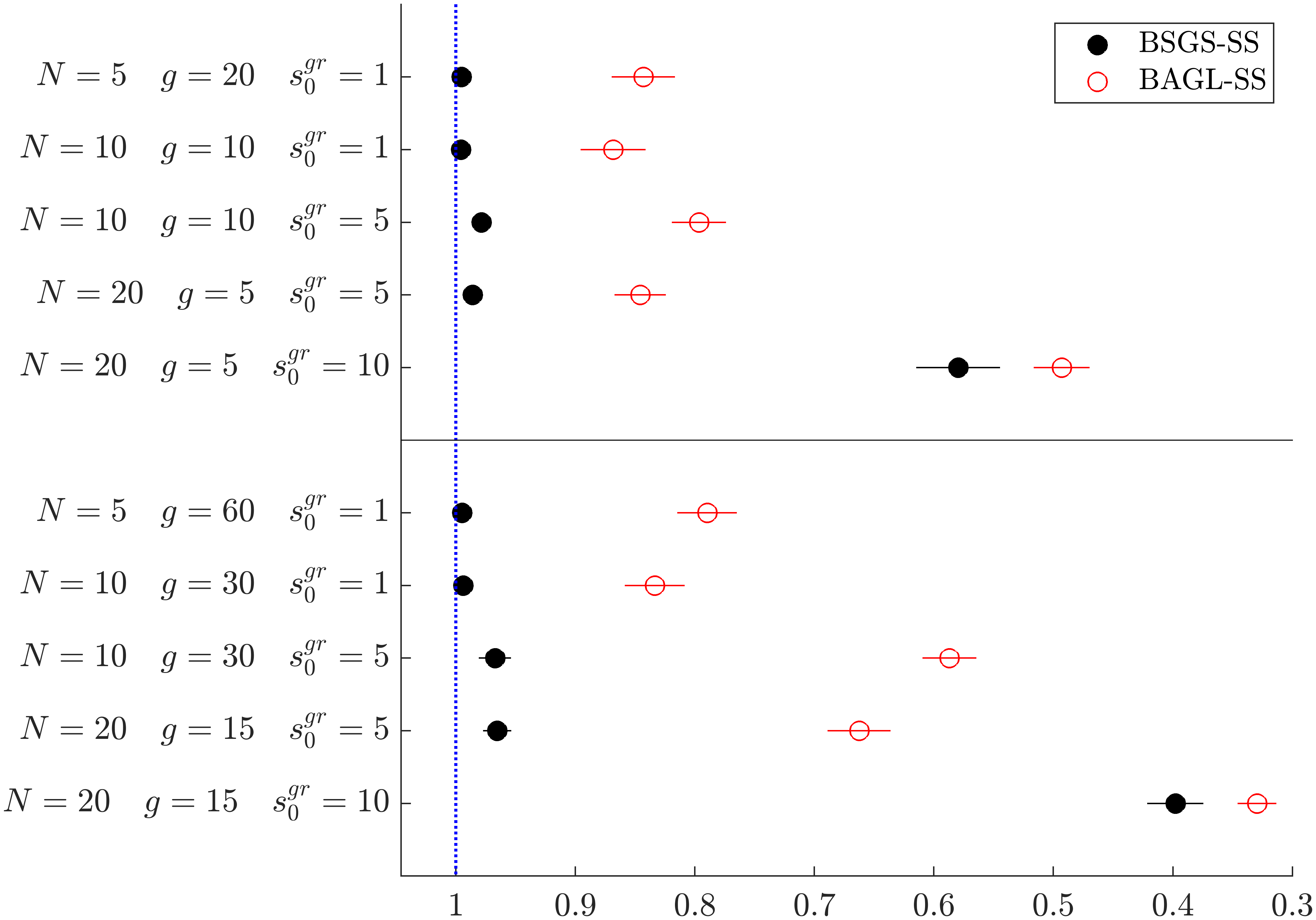}
    \vspace{-0.5cm}
    \caption{$\rho_{\epsilon}=0.75$}
\end{subfigure}\\
\begin{subfigure}[b]{0.50\textwidth}
	\centering
		  \includegraphics[width=\linewidth]{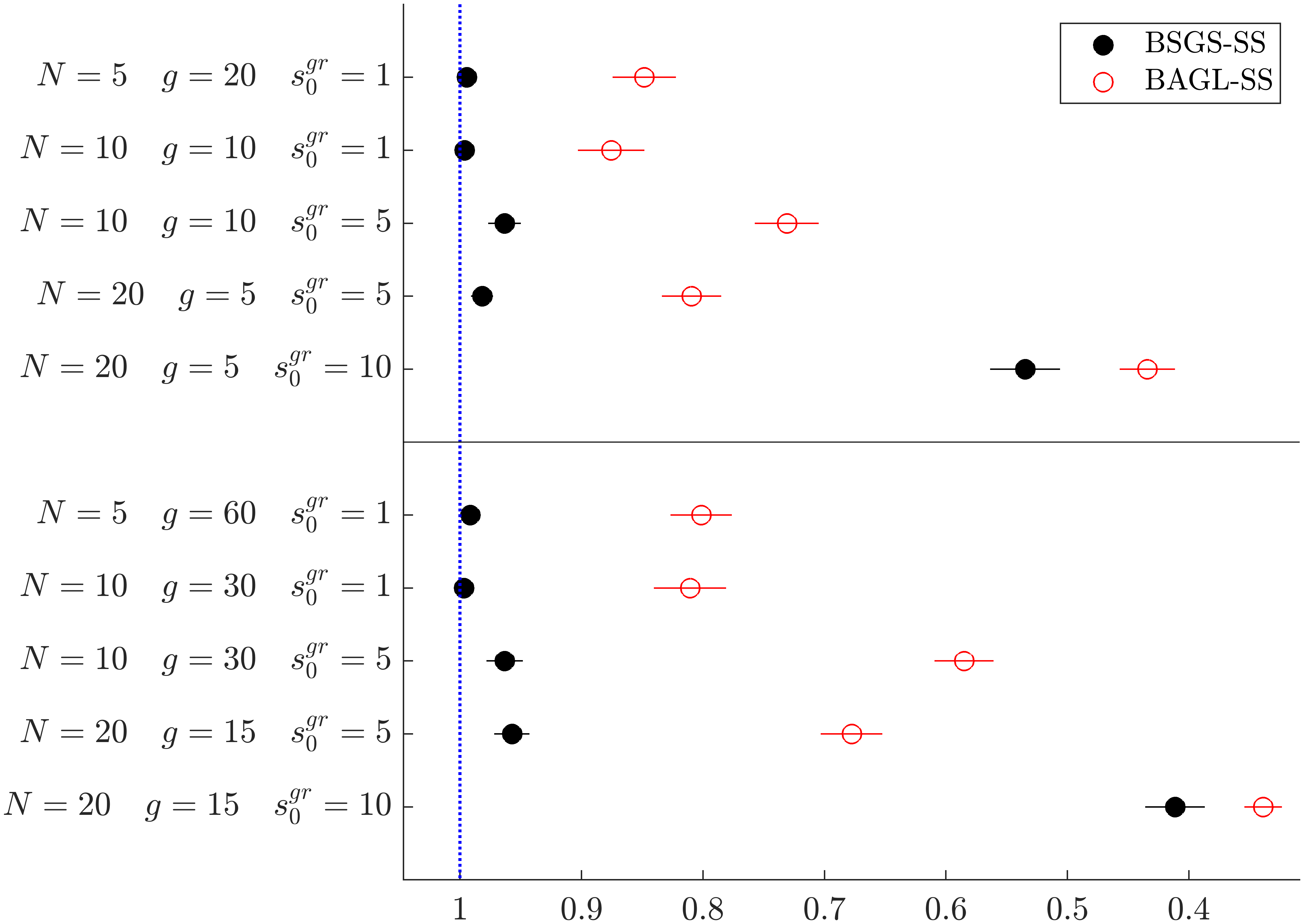}
    \vspace{-0.5cm}
    \caption{$\rho_{\epsilon}=0.75 ~\vert~ \mathcal{R}_{\epsilon}~\text{full}$}
\end{subfigure}
\begin{subfigure}[b]{0.50\textwidth}
	\centering
		  \includegraphics[width=\linewidth]{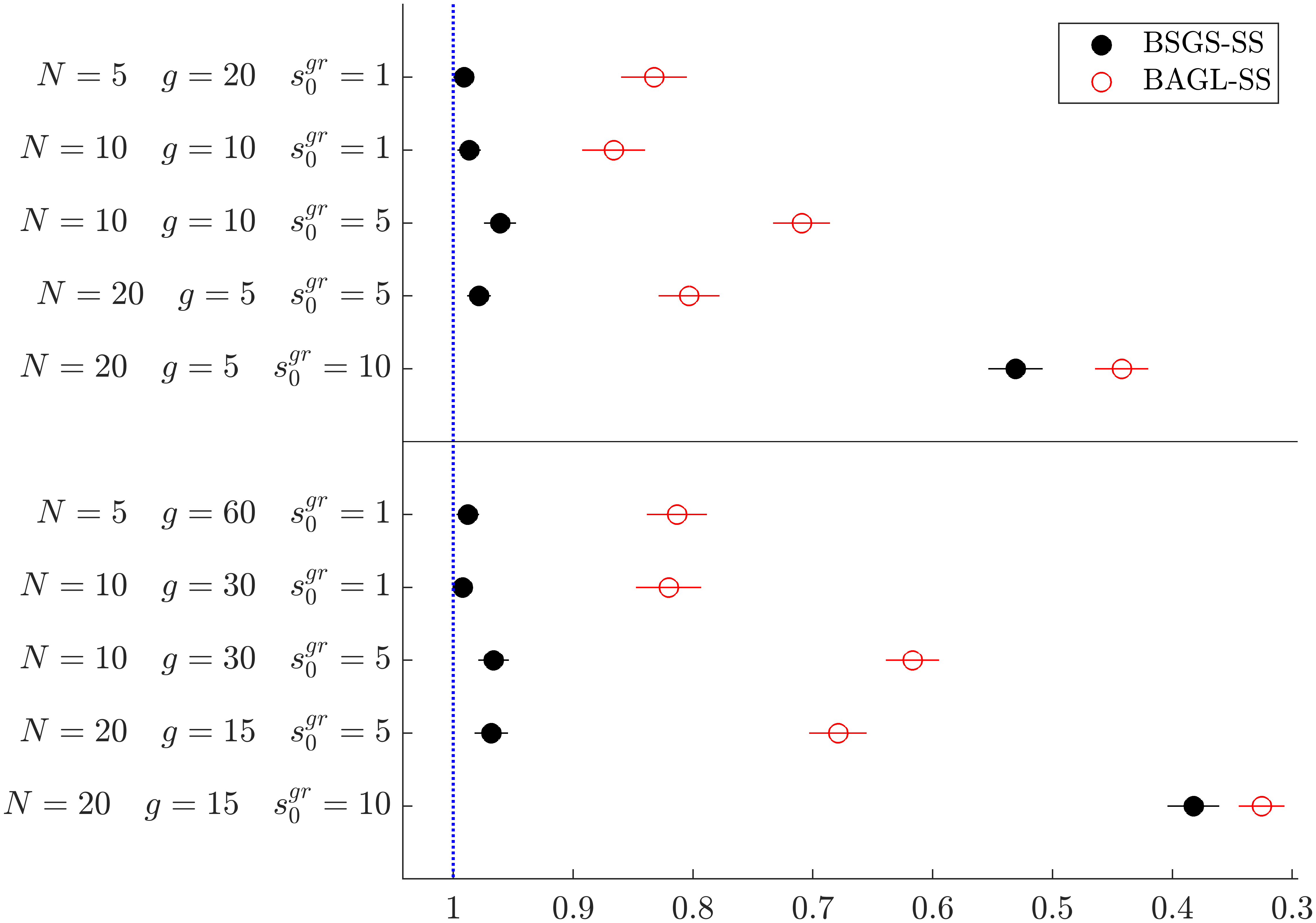}
        \vspace{-0.5cm}
    \caption{$\rho_{\epsilon}=0.75 ~\vert~ \mathcal{R}_{\epsilon}~\text{full} ~\vert~ \varepsilon_{t}\thicksim \text{Skew-}\mathcal{N}$}
\end{subfigure}
\caption*{\scriptsize Notes: The markers indicate the average Matthews correlation coefficient at the variable level (MCC$_g$) across all Monte Carlo iterations. Values close to one indicate a better overall quality of the identification of the true sparsity structure. The error bars denote $\pm$ 2 bootstrap standard errors of average values obtained from 200 Monte Carlo simulations.}
\end{figure}

As a final exercise, we compare the variable selection accuracy of the BSGS-SS prior and the BAGL-SS prior of \cite{MoglianiSimoni2021}. The latter relies on a single layer of sparsity (that is, sparsity at the group level only), and for this reason it is expected to be an appropriate prior only when dealing with specifications that are dense within each active group. Instead, when the true model is bi-level sparse, assuming a single-level group sparsity as in the BAGL-SS-prior, instead of bi-level group sparsity, does not ensure the posterior recovery of the correct sparsity pattern. 
This entails inaccurate selection and out-of-sample prediction. We investigate this issue numerically and the results are reported in Figure \ref{fig:MCCg}, which shows the MCC computed at the variable level (MCC$_g$) and across various DGPs. The simulation results confirm the theoretical results and strongly suggest a substantially lower selection accuracy for the BAGL-SS (MCC$_g<1$) compared to our BSGS-SS prior (MCC$_g$ close to one) when the true model is in fact bi-level sparse.

\subsection{Experiment 2: DGP with mixed-frequency data}\label{sec:MCsim_Exp2}
In this example, we consider a DGP with mixed-frequency data, where $N=\{50,100\}$ predictors are sampled at a higher frequency compared to the target variable.
Simulated data are obtained from the following process:
\begin{align}
y_{t} &= \alpha+\beta_{y} y_{t-1} + \sum_{j=1}^{N}\sum_{u=0}^{p_{x}}\psi_j\left(u\right)L^{\slfrac{u}{m}}x_{j,t-h}^{(m)} + \varepsilon_{t}, \label{MIDAS:simulation:1}\\[1.5ex]
x_{j,t}^{(m)} &= \rho_{x} x_{j,t-\slfrac{1}{m}}^{(m)} + \epsilon_{j,t},\nonumber
\end{align}
where $m=3$ (which is akin to a quarterly-monthly process) and $p_{x}=11$. We set the true weighting scheme $\psi_{j}\left(u\right)$ by relying on a three-parameters Beta function:
\begin{displaymath}
\psi_j\left(u\right)=\left(\frac{u+1}{p_{x}+1}\right)^{a-1}\left(1-\frac{u+1}{p_{x}+1}\right)^{b-1}\frac{\Gamma(a+b)}{\Gamma(a)\Gamma(b)}+c.
\end{displaymath}
%
%
%
We investigate four alternative weighting schemes that correspond to bell-shaped weights (DGP 1) with $(a,b,c)=(5,15,0)$, fast-decaying weights (DGP 2) with $(a,b,c)=(1,10,0)$, slow-decaying weights (DGP 3) with $(a,b,c)=(1,4,0)$, and flat weights (DGP 4) with $(a,b,c)=(1,1,0)$.\footnote{The weights are set to exactly zero for values of the Beta function $<1\text{e-04}$, and normalized to sum up to 1.}
Note that the same weighting scheme applies to all the predictors entering the DGPs.

We define $\varepsilon_{t}$ and $\boldsymbol\epsilon_{t} := (\epsilon_{1,t},\dots, \epsilon_{N,t})^{\prime}$ as i.i.d. draws from the normal multivariate distribution in \eqref{sim:1:errors}, where $\boldsymbol\Sigma_{\epsilon}=\mathcal{S}_{\epsilon}\mathcal{R}_{\epsilon}\mathcal{S}_{\epsilon}$, with $\mathcal{S}_{\epsilon}$ a diagonal matrix with elements $\sigma_{\epsilon}$ and $\mathcal{R}_{\epsilon}$ a Toeplitz correlation matrix with diagonal elements equal to one and off-diagonal elements $\rho_{\epsilon}^{|j-j^{\prime}|}$ for all $j\neq j^{\prime}$. We set $\rho_{\epsilon}=0.50$, $\alpha=0.50$, $\sigma=0.50$, $\rho_{x}=0.90$. The active set has cardinality $s_{0}^{gr}=\{5,10\}$ and the indices of active variables in $S_{0}^{gr}$ are fixed realizations of random draws without replacement from $\{1,\dots,N\}$. Conditional on these parameters, we calibrate $\sigma^2_{\epsilon}$ such that the noise-to-signal ratio of the mixed-frequency regression is $\text{NSR}=0.20$ for all the simulations. We assume $h=0$ (akin to a nowcasting model). The number of in- and out-of-sample observations is fixed at $T=200$ and $T_{oos}=50$, respectively.

\begin{table}[t!]
\footnotesize
\caption{Monte Carlo Experiment 2 - selection and density forecast accuracy}
\centering
\begin{tabular}{c|c|c|p{0.06\textwidth}<{\centering}p{0.06\textwidth}<{\centering}|p{0.06\textwidth}<{\centering}p{0.06\textwidth}<{\centering}|p{0.06\textwidth}<{\centering}p{0.06\textwidth}<{\centering}|p{0.06\textwidth}<{\centering}p{0.06\textwidth}<{\centering}}
\hline\hline
\multicolumn{3}{c}{} & \multicolumn{2}{c}{DGP 1} & \multicolumn{2}{c}{DGP 2} & \multicolumn{2}{c}{DGP 3} & \multicolumn{2}{c}{DGP 4}\\
\multicolumn{3}{c}{} & \multicolumn{2}{c}{bell-shaped} & \multicolumn{2}{c}{fast-decaying} & \multicolumn{2}{c}{slow-decaying} & \multicolumn{2}{c}{flat}\\
\hline
$N$ & $s_{0}^{gr}$ & Polynomial & TPR$_{N}$ & CRPS & TPR$_{N}$ & CRPS & TPR$_{N}$ & CRPS & TPR$_{N}$ & CRPS \\
\hline
\multirow{12}{*}{50} & \multirow{12}{*}{5} & \multirow{2}{*}{Unrestricted} & 99.8 & 0.71 & 100.0 & 0.67 & 99.2 & 0.75 & 70.5 & 0.88\\[-0.5ex]
			     &		                   & 	 		     & \scriptsize(0.2) & \scriptsize(0.01) & \scriptsize(0.0) & \scriptsize(0.01) & \scriptsize(0.5) & \scriptsize(0.01) & \scriptsize(3.3) & \scriptsize(0.01)  \\
			     &		                   & \multirow{2}{*}{Almon} & 84.3 & 0.75 & 81.8 & 0.75 & 85.0 & 0.77 & 94.8 & 0.77 \\[-0.5ex]
			     &		                   & 	 		     & \scriptsize(0.7) & \scriptsize(0.01) & \scriptsize(1.2) & \scriptsize(0.01) & \scriptsize(0.8) & \scriptsize(0.01) & \scriptsize(0.8) & \scriptsize(0.01) \\
			     &		                   & \multirow{2}{*}{Restr. Almon} & 99.5 & 0.70 & 87.5 & 0.71 & 98.8 & 0.71 & 100.0 & 0.79\\[-0.5ex]
			     &		                   & 	 		     & 	\scriptsize(0.3) & \scriptsize(0.01) & \scriptsize(0.7) & \scriptsize(0.01) & \scriptsize(0.4) & \scriptsize(0.01) & \scriptsize(0.0) & \scriptsize(0.01) \\
       			     &        			  & \multirow{2}{*}{Legendre} & 86.7 & 0.77 & 86.2 & 0.77 & 84.8 & 0.78 & 98.3 & 0.75 \\[-0.5ex]
			     &		                   & 	 		    & \scriptsize(0.7) & \scriptsize(0.01) & \scriptsize(0.9) & \scriptsize(0.01) & \scriptsize(1.1) & \scriptsize(0.01) & \scriptsize(0.5) & \scriptsize(0.01) \\
       			     &        			  & \multirow{2}{*}{Bernstein} & 99.3 & 0.71 & 97.2 & 0.67 & 100.0 & 0.70 & 90.8 & 0.84\\[-0.5ex]
			     &		                   & 	 		    & \scriptsize(0.3) & \scriptsize(0.01) & \scriptsize(0.6) & \scriptsize(0.01) & \scriptsize(0.0) & \scriptsize(0.01) & \scriptsize(2.2) & \scriptsize(0.01) \\
			     &        			  & \multirow{2}{*}{Chebychev T} & 87.5 & 0.76 & 81.7 & 0.76 & 84.0 & 0.78 & 98.7 & 0.75 \\[-0.5ex]
			     &		                   & 	 		    & \scriptsize(0.7) & \scriptsize(0.01) & \scriptsize(1.2) & \scriptsize(0.01) & \scriptsize(1.3) & \scriptsize(0.01) & \scriptsize(0.5) & \scriptsize(0.01) \\
\cline{1-11}			
\multirow{12}{*}{100} & \multirow{12}{*}{5} & \multirow{2}{*}{Unrestricted} & 99.3 & 0.72 & 99.2 & 0.69 & 98.2 & 0.74 & 42.2 & 0.94 \\[-0.5ex]
			     &		                   & 	 		     & \scriptsize(0.5) & \scriptsize(0.01) & \scriptsize(0.5) & \scriptsize(0.01) & \scriptsize(0.9) & \scriptsize(0.01) & \scriptsize(2.8) & \scriptsize(0.01)  \\
			     &		                   & \multirow{2}{*}{Almon} & 83.3 & 0.77 & 78.8 & 0.79 & 81.8 & 0.79 & 91.3 & 0.80 \\[-0.5ex]
			     &		                   & 	 		     & \scriptsize(0.3) & \scriptsize(0.01) & \scriptsize(1.5) & \scriptsize(0.01) & \scriptsize(1.1) & \scriptsize(0.01) & \scriptsize(1.1) & \scriptsize(0.01)  \\
			     &		                   & \multirow{2}{*}{Restr. Almon} & 97.3 & 0.71 & 83.5 & 0.74 & 94.8 & 0.71 & 93.7 & 0.83 \\[-0.5ex]
			     &		                   & 	 		     & \scriptsize(0.6) & \scriptsize(0.01) & \scriptsize(0.2) & \scriptsize(0.01) & \scriptsize(0.8) & \scriptsize(0.01) & \scriptsize(1.4) & \scriptsize(0.01) \\
       			     &        			   & \multirow{2}{*}{Legendre} & 80.2 & 0.79 & 82.7 & 0.81 & 78.3 & 0.81 & 91.0 & 0.78 \\[-0.5ex]
			     &		                   & 	 		     & \scriptsize(1.2) & \scriptsize(0.01) & \scriptsize(1.5) & \scriptsize(0.01) & \scriptsize(1.7) & \scriptsize(0.01) & \scriptsize(2.1) & \scriptsize(0.01) \\
       			     &        			   & \multirow{2}{*}{Bernstein} & 98.7 & 0.72 & 94.5 & 0.69 & 99.3 & 0.70 & 66.8 & 0.89 \\[-0.5ex]
			     &		                   & 	 		     & \scriptsize(0.6) & \scriptsize(0.01) & \scriptsize(0.8) & \scriptsize(0.01) & \scriptsize(0.6) & \scriptsize(0.01) & \scriptsize(3.3) & \scriptsize(0.01) \\
			     &        			   & \multirow{2}{*}{Chebychev T} & 80.3 & 0.79 & 81.7 & 0.79 & 78.7 & 0.81 & 91.7 & 0.78 \\[-0.5ex]
			     &		                   & 	 		     & \scriptsize(1.3) & \scriptsize(0.01) & \scriptsize(1.0) & \scriptsize(0.01) & \scriptsize(1.6) & \scriptsize(0.01) & \scriptsize(1.8) & \scriptsize(0.01) \\
\cline{1-11}			
\multirow{12}{*}{100} & \multirow{12}{*}{10} & \multirow{2}{*}{Unrestricted} & 23.3 & 0.96 & 27.9 & 0.93 & 16.1 & 0.98 & 10.1 & 0.99 \\[-0.5ex]
			     &		                   & 	 		     & \scriptsize(2.1) & \scriptsize(0.01) & \scriptsize(2.7) & \scriptsize(0.01) & \scriptsize(1.3) & \scriptsize(0.01) & \scriptsize(0.3) & \scriptsize(0.00)  \\
			     &		                   & \multirow{2}{*}{Almon} & 20.8 & 0.98 & 10.6 & 0.99 & 21.7 & 0.98 & 25.5 & 0.96 \\[-0.5ex]
			     &		                   & 	 		     & \scriptsize(2.6) & \scriptsize(0.01) & \scriptsize(0.7) & \scriptsize(0.00) & \scriptsize(2.6) & \scriptsize(0.01) & \scriptsize(3.1) & \scriptsize(0.01)  \\
			     &		                   & \multirow{2}{*}{Restr. Almon} & 88.2 & 0.77 & 89.3 & 0.76 & 87.4 & 0.78 & 17.7 & 0.97 \\[-0.5ex]
			     &		                   & 	 		     & \scriptsize(2.0) & \scriptsize(0.01) & \scriptsize(1.1) & \scriptsize(0.01) & \scriptsize(2.3) & \scriptsize(0.01) & \scriptsize(1.0) & \scriptsize(0.00) \\
       			     &        			   & \multirow{2}{*}{Legendre} & 16.5 & 0.98 & 14.2 & 0.98 & 18.6 & 0.97 & 16.2 & 0.98 \\[-0.5ex]
			     &		                   & 	 		     & \scriptsize(1.7) & \scriptsize(0.01) & \scriptsize(0.6) & \scriptsize(0.00) & \scriptsize(2.1) & \scriptsize(0.01) & \scriptsize(2.2) & \scriptsize(0.01) \\
       			     &        			   & \multirow{2}{*}{Bernstein} & 17.1 & 0.98 & 62.0 & 0.80 & 20.4 & 0.95 & 10.5 & 0.99 \\[-0.5ex]
			     &		                   & 	 		     & \scriptsize(0.9) & \scriptsize(0.01) & \scriptsize(3.6) & \scriptsize(0.02) & \scriptsize(2.1) & \scriptsize(0.01) & \scriptsize(0.4) & \scriptsize(0.00) \\
			     &        			   & \multirow{2}{*}{Chebychev T} & 19.1 & 0.97 & 10.8 & 0.99 & 19.1 & 0.98 & 13.1 & 0.98 \\[-0.5ex]
			     &		                   & 	 		     & \scriptsize(2.0) & \scriptsize(0.01) & \scriptsize(0.4) & \scriptsize(0.00) & \scriptsize(2.3) & \scriptsize(0.00) & \scriptsize(1.5) & \scriptsize(0.00) \\
\hline\hline
\multicolumn{11}{l}{\scriptsize \parbox[t]{14.4cm}{Notes: TPR$_N$ and CRPS denote respectively the True Positive Rate (at the group level) and the Continuously Ranked Probability Score, the latter in relative terms with respect to the AR(1) benchmark. Average values over 200 Monte Carlo simulations. Bootstrap standard errors of Monte Carlo averages in parentheses.}}
\end{tabular}
\label{t:mc_results4}
\end{table}

In the simulations, we consider and evaluate a set of alternative approximating functions for the true weighting function $\psi_j\left(u\right)$: the U-MIDAS \citep{Foroni2015}, Almon lag polynomials (restricted and unrestricted; \citealp{MoglianiSimoni2021}), and orthogonal lag polynomials (Legendre, Chebychev (first kind), and Bernstein orthogonal polynomials). 


Selection and predictive accuracy performance of our BSGS-SS procedure are reported in Table \ref{t:mc_results4}, where we report the True Positive Rate at the group level (TPR$_N$) and the CRPS. The results point to a number of interesting features. First, among the lag polynomials considered, the best results are provided by the unrestricted, the restricted Almon and the Bernstein polynomials. Second, consistently with the theory, the results for the best-performing polynomials are overall unaltered by the increase in the total number of high-frequency predictors. However, the performance is substantially affected by the increase in the number of true active variables. In this case, the restricted Almon performs surprisingly better than the other polynomials, which show, for some DGPs, very low selection and predictive accuracy. Third, the ranking of the best-performing polynomials may depend, at least in part, on the shape of the underlying true weighting function. For instance, the restricted Almon seems well suited for bell-shaped and slow-decaying weights, but somewhat less for fast-decaying weights, while unrestricted and Bernstein polynomials can perform fairly well irrespective of the underlying weighting structure.\footnote{However, we show that the U-MIDAS provides systematically less accurate parameter estimates compared to restricted Almon and Bernstein polynomials. These results are illustrated by the MSE in Figure A.2 in the Online Appendix.}
The performance of Legendre and Chebychev polynomials improves substantially under flat weights. This weighting structure is nevertheless less likely to describe the actual temporal aggregation for most economic data.

\begin{figure}[!t]
\caption{Monte Carlo Experiment 2 - relative CRPS score of our BSGS-SS prior with respect to the BAGL-SS and BSGL priors}
\begin{center}
\includegraphics[scale=0.65]{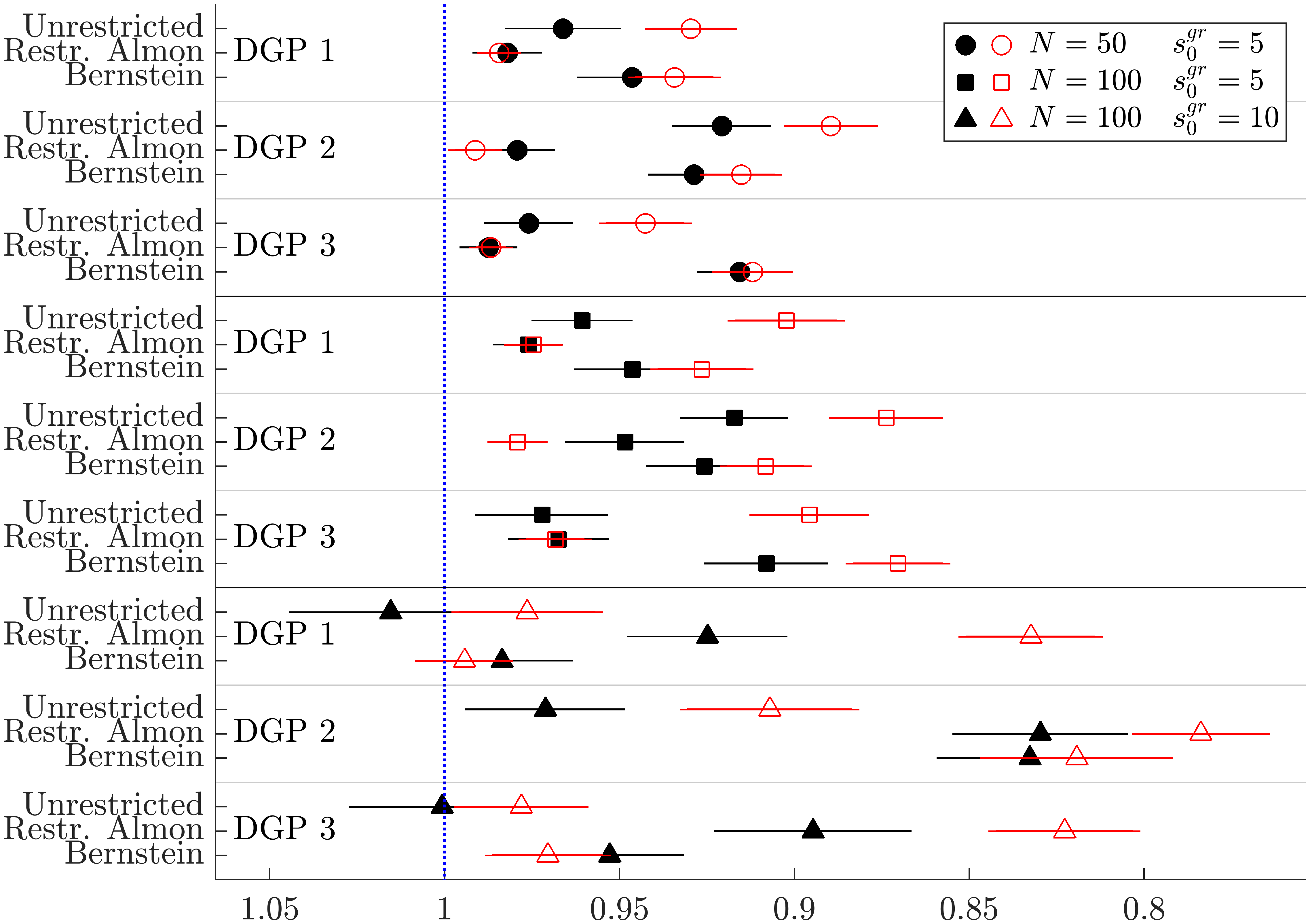}
\end{center}
\scriptsize \parbox[t]{14.4cm}{Notes: the markers denote the average relative CRPS score of our BSGS-SS prior with respect to the BAGL-SS (full markers) and BSGL (empty markers) priors. The error bars denote $\pm$ 2 bootstrap standard errors of average values obtained from 200 Monte Carlo simulations.}
\label{fig:RelativeBMIDASsim_CRPS}
\end{figure}

Finally, we compare again the predictive accuracy of our BSGS-SS prior with respect to the BSGL and BAGL-SS  benchmarks. Unlike Experiment 1, here we do not control directly for the degree of sparsity (in the coefficients of the approximating functions) within the groups, but this arises indirectly from the specification of DGPs 1 to 4 considered. The results reported in Figure \ref{fig:RelativeBMIDASsim_CRPS} suggest that the BSGS-SS prior provides a better predictive performance also in the mixed-frequency framework: the relative CRPS (with respect to the benchmarks) is on average less than one across the considered DGPs and polynomials (the best performing ones according to the results in Table \ref{t:mc_results4}), pointing to an average gain hovering around 5\% to 10\% overall.



\section{Empirical application: nowcasting US GDP in a mixed-frequency framework}\label{sec:emp_app}
We use the proposed prior for a nowcasting exercise of US GDP in the mixed-frequency framework \eqref{MIDAS:simulation:1} of Experiment 2, 
where $y_{t}=400\log(Y_{t}/Y_{t-1})$ is the annualized quarterly growth rate of GDP, and $\mathbf{x}_{t}$ is a vector of $122$ macroeconomic series sampled at monthly frequency and extracted from the FRED-MD database \citep{McCracken2016}. The data sample starts in 1980Q1, while the pseudo out-of-sample analysis spans 2013Q1 to 2022Q4. Estimates are carried-out recursively using a rolling window of $T=132$ quarterly observations, and $h$-step-ahead posterior predictive densities are generated from \eqref{eq:pred_dens} through a direct forecast approach. We consider 3 nowcasting horizons ($h=0,1/3,2/3$) and the two lag polynomials that provide overall better results in the simulation analysis reported in Section \ref{sec:MCsim_Exp2} -- namely the restricted Almon and the orthonormal Bernstein polynomials. Further, as the considered data sample spans periods characterised by strong and uneven macroeconomic fluctuations (permanent and transitory), such as the Great Moderation, the Great Recession, and the Covid crisis, we also allow for time-varying volatility with heavy tails and occasional outliers in the regression errors by using the model in Section \ref{sec:SVOtARMA}. Finally, we do not take into account real-time issues (ragged/jagged-edge data and revisions) and the data used for the analysis reflect the latest vintages available at the time of writing.\footnote{The GDP series used in the analysis was downloaded from FRED-MD database on July 2023 and we consider the 2023-06 vintage. We exclude from the analysis 5 series presenting sampling issues over the time span considered for the analysis (namely, new orders for consumer goods, non-borrowed reserves of depository institutions, 3-month AA financial commercial paper rate, 3-month commercial paper minus fed funds, consumer sentiment index).}\\
\indent We consider several modelling strategies to exploit our bi-level sparsity prior approach. First, we estimate the forecasting model \eqref{MIDAS:simulation:1} on the whole set of 122 indicators. 
Then, since the simulation results in Section \ref{sec:MCsim} point to a substantial deterioration of the selection and prediction performance in extreme sparse settings where $Ng\gg T$ then, we also estimate a separate model for each of the 8 groups of indicators partitioned as in \cite{McCracken2016}.\footnote{The groups encompass \textit{i)} output and income, \textit{ii)} labour market, \textit{iii)} housing, \textit{iv)} consumption, orders, and inventories, \textit{v)} money and credit, \textit{vi)} interest and exchange rates, \textit{vii)} prices, and \textit{viii)} stock market. The number of indicators included in each group ranges between 5 and 31.}
Summing up, we estimate a large set of alternative specifications, according to the 2 lag polynomials (Almon and Bernstein), the 5 volatility process (homoskedastic, SV, SV with Student-$t$ shocks, SV with outliers, SV with Student-$t$ shocks and outliers), and the 2 partition strategies (whole dataset \textit{vs} 8 groups). 
For each of these specifications we compute density forecasts and we combine them along several dimensions (across the 8 groups, the volatility process, and the lag polynomials, see Online Appendix A.4 for more details).
Density forecasts combination is carried out using the optimal prediction pool proposed by \citet{Geweke2011}, which relies on log-scores. 

\begin{table}[t!]
\footnotesize
\caption{Empirical application: nowcasting US GDP - CRPS}
\centering
\begin{tabular}{l |ccc|ccc}
\hline\hline
 & \multicolumn{3}{c|}{BSGS-SS} & \multicolumn{3}{c}{BSGL} \\
\hline
 &  $h=0$ & $h=1/3$ & $h=2/3$ & $h=0$ & $h=1/3$ & $h=2/3$ \\
\hline
\cline{1-7}
\multirow{2}{*}{Groups - Almon} & 0.78 & 0.81 & 0.76 & 0.82 & 0.81 & 0.77 \\[-0.5ex]
			& \scriptsize(0.02) & \scriptsize(0.02) & \scriptsize(0.00) & \scriptsize(0.16) & \scriptsize(0.02) & \scriptsize(0.00) \\
\multirow{2}{*}{Groups - Bernstein} & \textbf{0.70} & 0.70 & 0.79 & 0.77 & 0.73 & 0.76 \\[-0.5ex]
			      & \scriptsize(0.01) & \scriptsize(0.00) & \scriptsize(0.02) & \scriptsize(0.04) & \scriptsize(0.00) & \scriptsize(0.00) \\
\multirow{2}{*}{Groups - all} & 0.71 & \textbf{0.69} & 0.77 & 0.77 & 0.74 & 0.78 \\[-0.5ex]
		   & \scriptsize(0.01) & \scriptsize(0.00) & \scriptsize(0.01) & \scriptsize(0.04) &	\scriptsize(0.00) & \scriptsize(0.01) \\
\multirow{2}{*}{Whole dataset - Almon} & 0.71 & 0.81 & \textbf{0.75} & 0.81 & 0.77 & 0.79 \\[-0.5ex]
				    & \scriptsize(0.01) & \scriptsize(0.06) & \scriptsize(0.00) & \scriptsize(0.23) & \scriptsize(0.01) & \scriptsize(0.00) \\
\multirow{2}{*}{Whole dataset - Bernstein} & 0.71 & 0.75 & 0.83 & 0.85 & 0.89 & 0.86 \\[-0.5ex]
					& \scriptsize(0.01) & \scriptsize(0.01) & \scriptsize(0.03) & \scriptsize(0.27) & \scriptsize(0.35) & \scriptsize(0.07) \\
\multirow{2}{*}{Whole dataset - all} & 0.70 & 0.78 & 0.76 & 0.85 & 0.84 & 0.87 \\[-0.5ex]
			      & \scriptsize(0.01) & \scriptsize(0.03) & \scriptsize(0.00) & \scriptsize(0.30) & \scriptsize(0.05) & \scriptsize(0.08) \\
\hline\hline
\multicolumn{7}{l}{\scriptsize \parbox[t]{12.9cm}{Notes: the Table displays the Continuously Ranked Probability Score (CRPS), in relative terms with respect to the AR(1) benchmark. BSGS-SS denotes the proposed bi-level sparsity prior. BSGL denotes the Bayesian Sparse Group Lasso prior \citep{Xu2015}. $p$-values of a test of unconditional predictive accuracy \citep{Giacomini2006} with respect to the AR(1) benchmark in parentheses. 
Bold numbers denote the best outcomes across horizons and models.}}
\end{tabular}
\label{t:emp_results}
\end{table}

The combined density forecasts are then evaluated by the means of the average CRPS. The relative scores with respect to the AR(1) benchmark are reported in Table \ref{t:emp_results}. 
Three main findings can be outlined. First, Bernstein polynomials provide accurate density forecasts for $h=0$ and $h=1/3$, but the restricted Almon performs best at $h=2/3$. This outcome can be due to the way the different polynomials best describe the underlying weighting structure at each horizon. However, pooling these lag polynomials does not seem to provide here a systematic optimal strategy. Second, there is no clear-cut evidence in favour of a particular partition strategy. For $h=0$ and $h=2/3$, models including the whole set of indicators seem to perform better, or very closely, to those based on groups. Conversely, pooled forecasts from models based on group partitions tend to outperform for $h=1/3$. Overall, and consistently with the simulation results reported in Section \ref{sec:MCsim_Exp2}, this outcome suggests that the proposed procedure is robust to the presence of a large number of predictors. Finally, when compared to the BSGL prior (right hand side of Table \ref{t:emp_results}), the results point to a considerable outperformance of our prior and substantial predictive gains for some horizons and pooling strategies.


\section{Concluding remarks}\label{sec:conclusions}

This paper constructs optimal density forecasts for macroeconomic indicators in high-dimensional settings where the covariates present a known group structure. Such a structure arises, for instance, because variables in datasets are organised in groups or because of the flexible modelling of the relationship between the target variable and the set of covariates. The group-structure is important in order to reduce the dimensionality and we propose an optimal way to exploit it via the specification of a convenient prior distribution. By working under the assumption of bi-level sparsity, \textit{i.e.} among and within groups, our procedure is able to detect the driving factors if only some regressors are relevant for forecast.

Density forecasts are constructed from the posterior predictive density, which we prove to present optimal asymptotic properties and to outperform many competitors in finite sample. We also provide parameter recovery and point forecasts, which we prove to be minimax-optimal for our procedure under some mild conditions. Unlike alternative approaches, our procedure does not require orthogonality among covariates belonging to different groups. 
Finally, we show that a more general error structure, such as stochastic volatility and ARMA, can be accommodated by only slightly modifying the proposed prior.


\vspace{-0.2cm}
  \paragraph{Acknowledgements.}
  The second author gratefully acknowledges financial support from Hi!Paris, and ANR-21-CE26-0003. The usual disclaimer applies. The views expressed in this paper are those of the authors and do not necessarily reflect those of the Banque de France or the Eurosystem.



\vspace{-0.2cm}
  \paragraph{Online Appendix.}\label{App:supplementary}
  It contains: additional elements about the prior, the Monte Carlo experiments and the application, guidelines for setting hyperparameters, the MCMC algorithm for the extended model with ARMA errors and stochastic volatility, additional theoretical results and all the proofs.

\vspace{-0.2cm}
\bibliographystyle{elsarticle-harv}
\onehalfspacing
\small
\bibliography{Manuscript}

\end{document}